\documentclass[iop,apj,revtex4,tighten]{emulateapj}

\usepackage[english]{babel}
\usepackage[utf8x]{inputenc}
\usepackage{amsmath}
\usepackage{graphicx}
\usepackage{subfigure}
\def\simgt{\lower 2pt \hbox{$\, \buildrel {\scriptstyle >}\over{\scriptstyle \sim}\,$}}
\def\simlt{\lower 2pt \hbox{$\, \buildrel {\scriptstyle <}\over{\scriptstyle \sim}\,$}}

\begin{document}

\title{The role of galactic cold gas in low-level supermassive black
  hole activity}

\author{Erik D.~Alfvin$^1$, Brendan P.~Miller$^2$, Martha
  P.~Haynes$^3$, Elena Gallo$^4$, Riccardo Giovanelli$^3$, \\ Rebecca
  A.~Koopmann$^5$, Edmund Hodges-Kluck$^4$, John M.~Cannon$^1$}

\footnotetext[1]{Department of Physics and Astronomy, Macalester College,
  Saint Paul, MN 55105, USA}

\footnotetext[2]{Department of Chemistry and Physical Sciences, The
  College of St. Scholastica, Duluth, MN 55811, USA}

\footnotetext[3]{Center for Radiophysics and Space Research, Space
  Sciences Building, Cornell University, Ithaca, NY 14853, USA}

\footnotetext[4]{Department of Astronomy, University of Michigan, Ann
  Arbor, MI 48109, USA}

\footnotetext[5]{Department of Physics and Astronomy, Union College,
  Schenectady, NY 12308, USA}

\begin{abstract}
The nature of the relationship between low-level supermassive black
hole (SMBH) activity and galactic cold gas, if any, is currently
unclear. Here, we test whether central black holes may feed at higher
rates in gas-rich galaxies, probing SMBH activity well below the
active regime down to Eddington ratios of $\sim10^{-7}$. We use a
combination of radio data from the ALFALFA survey and from the
literature, along with archival X-ray flux measurements from the {\it
  Chandra\/} X-ray observatory, to investigate this potential
relationship. We construct a sample of 129 late-type galaxies, with
$M_{\rm B}<-18$ out to 50 Mpc, that have both HI masses and sensitive
X-ray coverage. Of these, 75 host a nuclear X-ray source, a {\bf 58\%}
detection fraction. There is a highly significant correlation between
nuclear X-ray luminosity $L_{\rm X}$ and galaxy stellar mass $M_{\rm
  star}$ with a slope of 1.7$\pm$0.3, and a tentative correlation
(significant at the 2.8$\sigma$ level) between $L_{\rm X}$ and HI gas
mass $M_{\rm HI}$. However, a joint fit to $L_{\rm X}$ as a function
of both $M_{\rm star}$ and $M_{\rm HI}$ finds no significant
dependence on $M_{\rm HI}$ (slope $0.1\pm0.3$), and similarly the
residuals of $L_{\rm X}-L_{\rm X}(M_{\rm star})$ show no trend with
$M_{\rm HI}$; the apparent correlation between $L_{\rm X}$ and $M_{\rm
  HI}$ seems to be entirely driven by $M_{\rm star}$. We demonstrate
quantitatively that these results are robust against X-ray binary
contamination. We conclude that the galaxy-wide cold gas content in
these spirals does not strongly correlate with their low-level
supermassive black hole activity, and suggest fueling is a localized
process.
\end{abstract}


\section{Introduction} 

Nearly every moderately massive galaxy (i.e., $M_{\rm star} > 10^{10}
M_{\odot}$) appears to contain a supermassive black hole (SMBH) at its
center, with $M_{\rm BH}$ correlating with the bulge stellar velocity
dispersion $\sigma$, optical luminosity, or stellar mass
\citep[e.g.,][and references
  therein]{2009ApJ...698..198G,2013ApJ...764..184M}. Some active
galactic nuclei are rapidly accreting gas at near Eddington rates, but
most SMBHs in the local universe are weakly accreting or
quasi-quiescent, like our own Milky Way. The galactic processes that
drive SMBH accretion are complex and gas movement within disks is not
yet well understood. However, black hole mass, which necessarily
increases through accretion, is well studied across different types of
galaxies. Compared to bulge-dominated early-type galaxies, where the
bulge accounts for nearly all of the luminosity and stellar mass,
late-type galaxies follow a $M_{\rm BH}-\sigma$ relation with a
similar slope (black hole mass positively correlating with higher
velocity dispersion) but a lower intercept. The SMBHs hosted in
spirals are therefore on average less massive by a factor of $\sim$2
than in ellipticals at a given $\sigma$ \citep{2013ApJ...764..184M},
although some studies do find spirals and ellipticals show consistent
trends (Graham \& Scott (2013). Late-type galaxies also offer some
disk tracers of SMBH mass, such as spiral arm pitch angle
\citep[e.g.,][and references therein]{2014ApJ...789..124D}, and it is
now clear that there are bulgeless galaxies with SMBHs
\citep{2011Natur.470...66R,2013MNRAS.429.2199S,2014ApJ...782...22B,2014ApJ...784..113S}. SMBH
masses do not appear to correlate with disk or pseudobulge absolute
magnitudes \citep{2011Natur.469..374K}, although see also discussion
in L{\"a}sker et al.~(2014). These factors suggest that black hole
growth could be tied to galactic gas, and not only to bulge
properties. The rate of major mergers is lower for late-type galaxies
than for early-type galaxies (spiral morphology can be destroyed in
major mergers, although gas rich mergers may preserve or regrow disks;
Hopkins et al.~2009), so they are less able to activate and
significantly grow their SMBHs through mergers.

The relationship between SMBH growth and activity, and host galaxy
evolution is complex (Shankar et al.~2009). Cold gas reservoirs could
fuel star formation and SMBH accretion (the latter perhaps indirectly,
from stellar winds) and we would observe a positive correlation
between SMBH accretion rate and cold gass mass.  On the other hand, if
SMBH accretion feedback heats and expels gas and thereby shuts down
star formation, we might observe a negative correlation. Or,
particularly in smaller galaxies with smaller SMBHs, timescale
mismatches and a limited energy budget may causally decouple these
processes \citep[e.g.,][and references
  therein]{2013ARA&A..51..511K}. Here, we explore the black hole
accretion - cold gas mass relationship through investigating how
nuclear X-ray luminosities are linked to total HI masses. This work
both complements and extends previous studies in that we investigate
all detectable SMBH activity in nearby, late-type galaxies rather than
only known active galactic nuclei (AGNs); indeed, the majority of
galaxies in our sample do not have optical indicators of nuclear
activity.

As also noted for nearby early-type galaxies, there is more than
sufficient (hot) gas near the central regions (Soria et al.~2006) to
produce observed levels of activity. That being said, it seems
plausible that additional (cold) gas, either directly available for
SMBH accretion or indirectly involved via star formation that then
adds mass loss from winds, would increase the accretion rate and
consequent X-ray emission. We might expect that galaxies with larger
total HI masses transport more of that cold gas to the central region,
and have more active SMBHs. The expected rates of mass accretion for
these late-type galaxies correspond to doubling times of tens of Gyr,
but addressing this question is nonetheless important for
understanding SMBH growth. A positive correlation between galaxy-wide
HI mass and nuclear X-ray emission might suggest that cold gas is
important for activity also in powerful AGNs. In this context SMBH
growth would not need to be self-regulated by feedback but rather
would depend on the transport properties (e.g., torque-limited
accretion; Angles-Alcazar et al.~2013).

Our HI masses are primarily drawn from the ALFALA survey. ALFALFA is a
blind HI survey using the Arecibo L-band Feed Array (ALFA) at the
Arecibo telescope to scan a portion of the
$0^{\circ}<\delta<36^{\circ}$ sky at frequencies surrounding the HI
1420 MHz line (1335--1435 MHz). The $\alpha$40 preliminary catalog
(40\% coverage complete) is presented in Haynes et al.~(2011). Here,
we use the new $\alpha$70 catalog\footnote{{\tt
    http://egg.astro.cornell.edu/alfalfa/data/}} (Haynes et al., in
preparation; results for 70\% coverage) to obtain HI masses for a
sample of local spiral galaxies. We supplement the ALFALFA data with
literature HI masses taken from the HyperLeda database\footnote{\tt
  http://leda.univ-lyon1.fr/} \citep{2014A&A...570A..13M}

For X-ray luminosities we use archival {\it Chandra\/} data. High
angular resolution {\it Chandra\/} X-ray observations provide a proven
method of identifying low-level supermassive black hole activity in
early-type galaxies (e.g., Zhang et al.~2009; Gallo et al.~2010;
Pellegrini 2010; Miller et al.~2012). In contrast to ellipticals,
which do not typically contain large amounts of cold gas (particularly
in clusters; e.g., Grossi et al.~2009 and references therein), spirals
are often gas-rich. A gas-rich environment provides a complicating
contamination possibility from high-mass X-ray binaries (which can be
present near sites of recent star formation), but here too {\it
  Chandra\/} generally has sufficient resolution to identify SMBH
activity (e.g., Mathur et al.~2010; O'Sullivan et al.~2014; Tzanavaris
et al.~2014). We develop a statistical technique to quantify potential
X-ray binary contamination and account for it in our linear regression
modeling. The X-ray coverage provides sensitivities pushing down to
luminosities relative to Eddington of $\sim10^{-7}$, well below the
formally active AGN regime.

\begin{figure}
\includegraphics[scale=0.365]{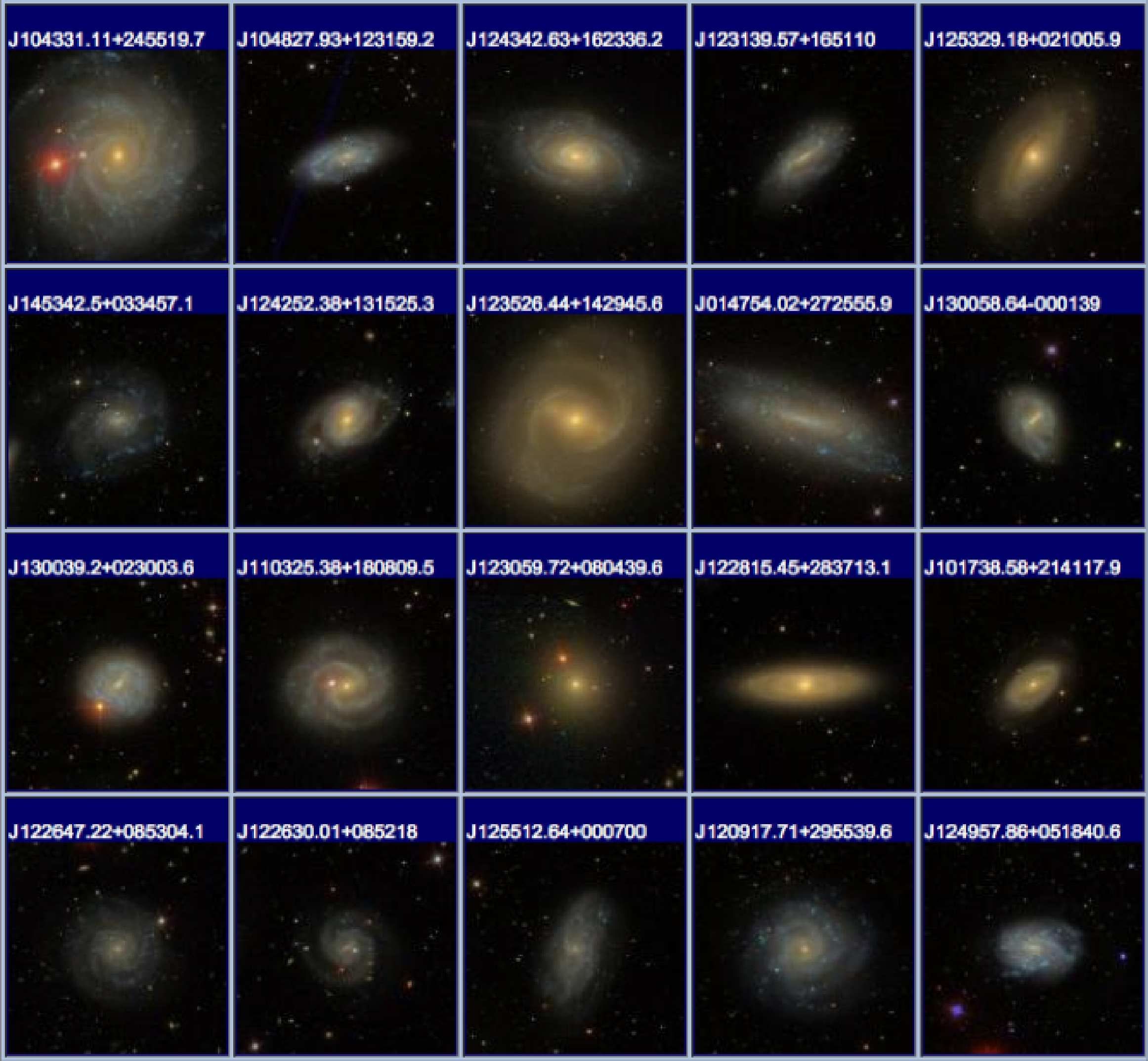} \figcaption{\small SDSS
  optical cutouts of 20 representative galaxies in our sample with
  ALFALFA HI measurements. All show clear spiral morphology.}
\end{figure}

This paper is organized as follows. Section 2 describes the sample
selection and calculation of HI masses and X-ray luminosities. Section
3 presents correlation analyses, controlling for potential X-ray
binary contamination, measurement uncertainties, and X-ray upper
limits. Section 4 discusses additional multiwavelength activity
indicators and the relationship between star formation rates, cold
gas, and SMBH activity, and potential future work.

\section{Sample selection and Characteristics}  

Here we describe the optical selection, measurement of HI masses, and
determination of X-ray coverage for our sample. SDSS cutouts verifying
spiral morphology for 20 representative galaxies with ALFALFA HI
measurements are provided in Figure~1. X-ray, optical, and ALFALFA HI
data are shown for reference for NGC 4501, one of the most luminous
spirals in our sample, in Figure~2.

\begin{figure}
\includegraphics[scale=0.45]{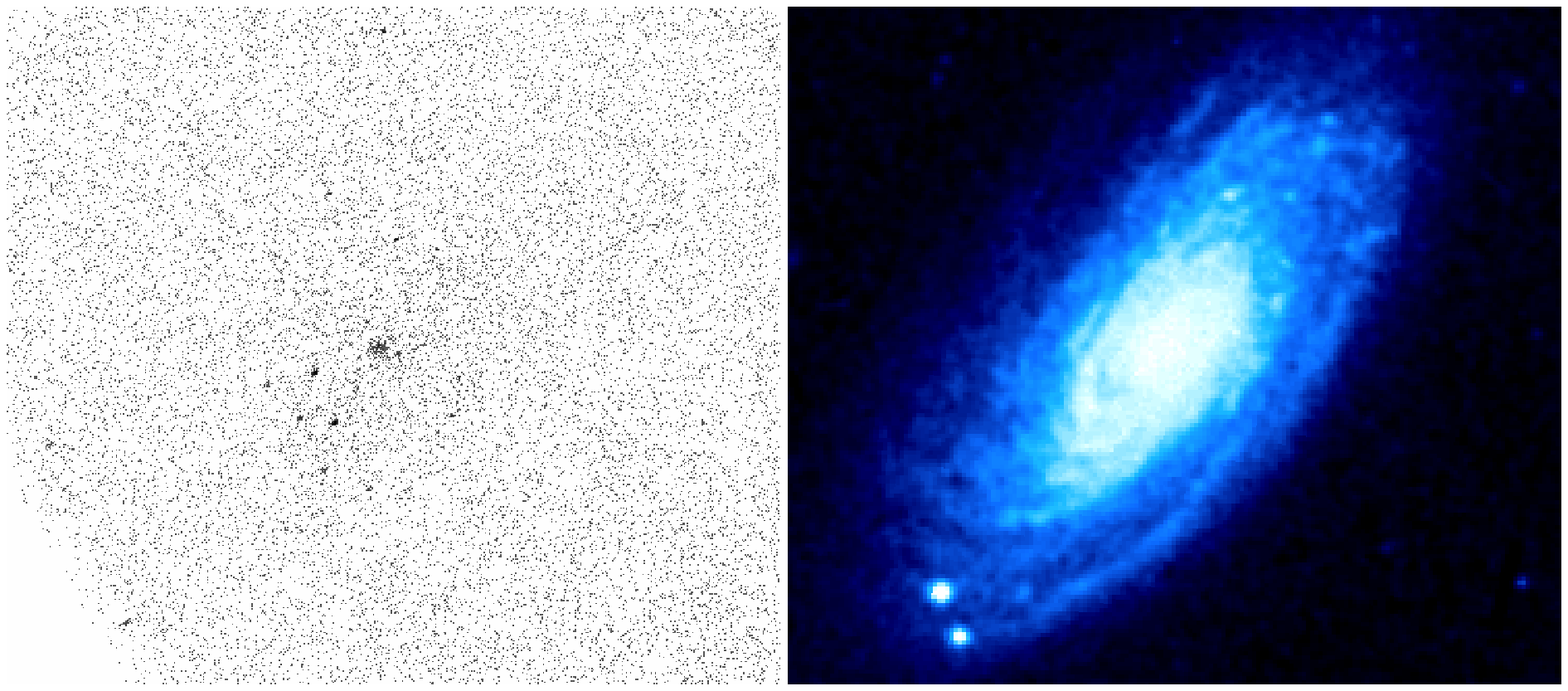}
\includegraphics[scale=0.4]{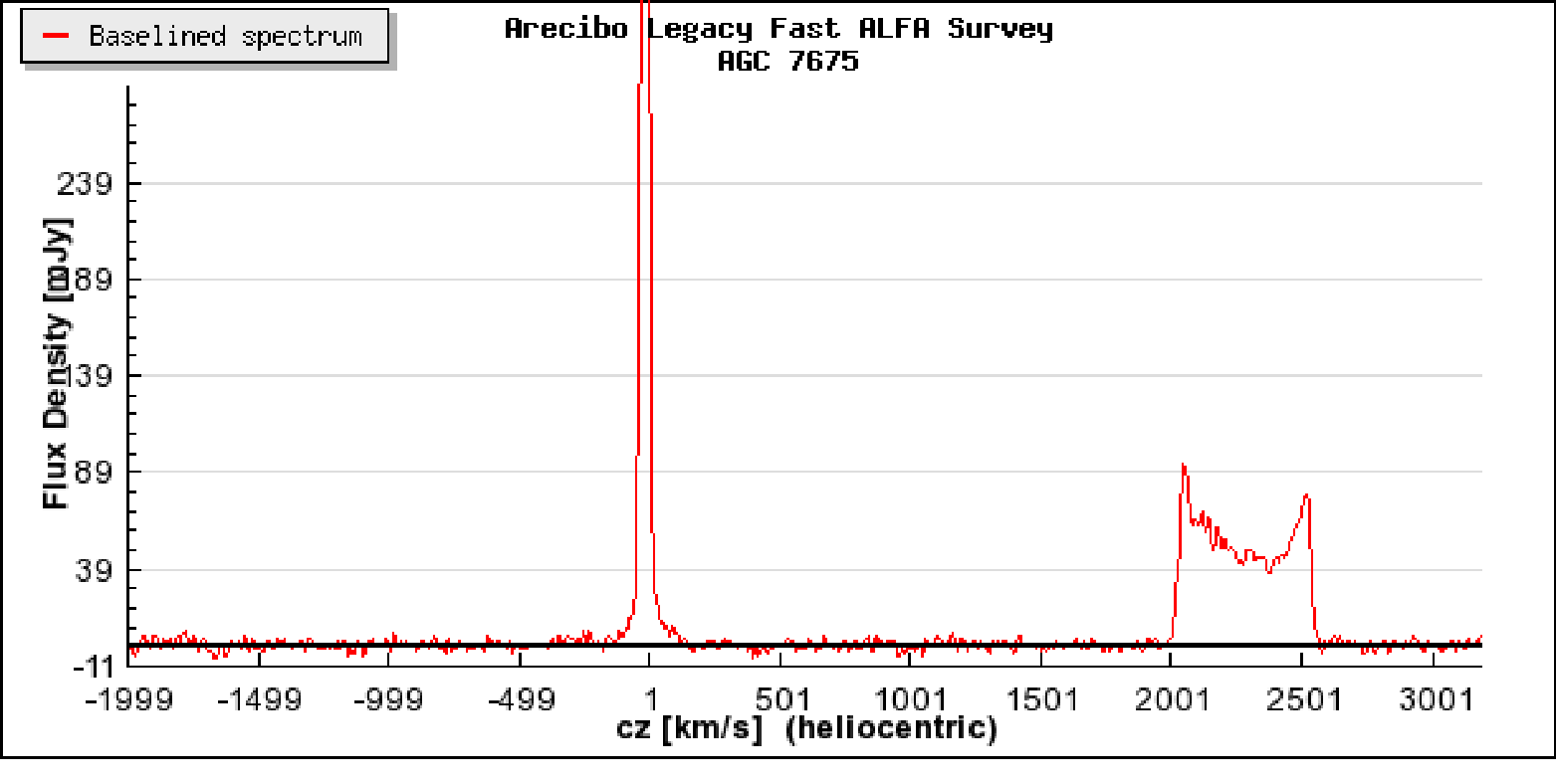} \figcaption{\small Left: Full
  band X-ray {\it Chandra\/} image of NGC 4501, showing nuclear X-ray
  source as well as a few XRBs. Right: DSS optical image at matched
  angular scale. Bottom: ALFALFA HI spectrum, showing double-peaked
  emission at the redshift of NGC 4501. The narrow emission line at
  $cz\sim0$ is due to the Milky Way.}
\end{figure}

\subsection{Optical selection of parent sample}

We selected a volume-limited parent sample of late-type galaxies from
the HyperLeda database. We required each source to be a spiral galaxy
with morphological type between $t=1$ (Sa) and $t=7$ (Sd) -- note that
this excludes S0 galaxies -- and imposed a luminosity limit of $M_{\rm
  B}<-18$. We additionally restrict consideration to spirals with
inclination between the polar axis and the line-of-sight $<70^{\circ}$
(i.e., not edge-on) to reduce the complicating effects of internal
extinction. We initially consider all late-type galaxies with a
distance modulus less than 34.8 (i.e., within $\simeq$90 Mpc).

The HyperLeda SQL query that we used to construct our parent sample
is: {\tt select objname, al2000, de2000, t, modz, mabs, bvtc, m21c
  where modz <= 34.8 and mabs <= -18 and t >= 1 and t <= 7 and incl <
  70 order by mabs}

There are 6491 galaxies that meet these criteria. Distances in Mpc for
the full parent sample are initially calculated from the redshift
distance modulus given in HyperLeda. After cross-matching to ALFALFA
and {\it Chandra\/} coverage and further refining selection criteria
(see below), for the final sample of 129 spirals with HI masses and
X-ray coverage we update the distances (and all associated
luminosities, absolute magnitudes, and masses) for the 126/132
galaxies that have non-redshift measurements available in the
Extragalactic Distance Database\footnote{{\tt
    http://edd.ifa.hawaii.edu/}} (EDD; Tully et al.~2009) listing of
preferred distances, which is restricted to $v<3000$~km~s$^{-1}$.

For the galaxies in the parent sample, we calculate the stellar mass
in solar units using the relations given in Bell et al.~(2003) as

\begin{equation*}
M_{\rm star} = 1.737(B-V)-0.942+0.4(B_{\odot}-B)
\end{equation*}

Here the absolute magnitude for the sun is taken to be
$B_{\odot}=5.515$.\footnote{{\tt
    http://sites.google.com/site/mamajeksstarnotes/}\\{\tt
    basic-astronomical-data-for-the-sun}} The $B-V$ color is the
difference between the $B$ and $V$ optical magnitudes corrected for
galactic extinction. For the parent sample, $M_{\rm star}$ is not
calculated for galaxies that lack $B-V$ colors in HyperLeda; for the
final sample of 129 spirals, missing $B-V$ colors (24\% of entries)
are set to the median 0.56 value.

\begin{figure*}
\includegraphics[scale=0.88]{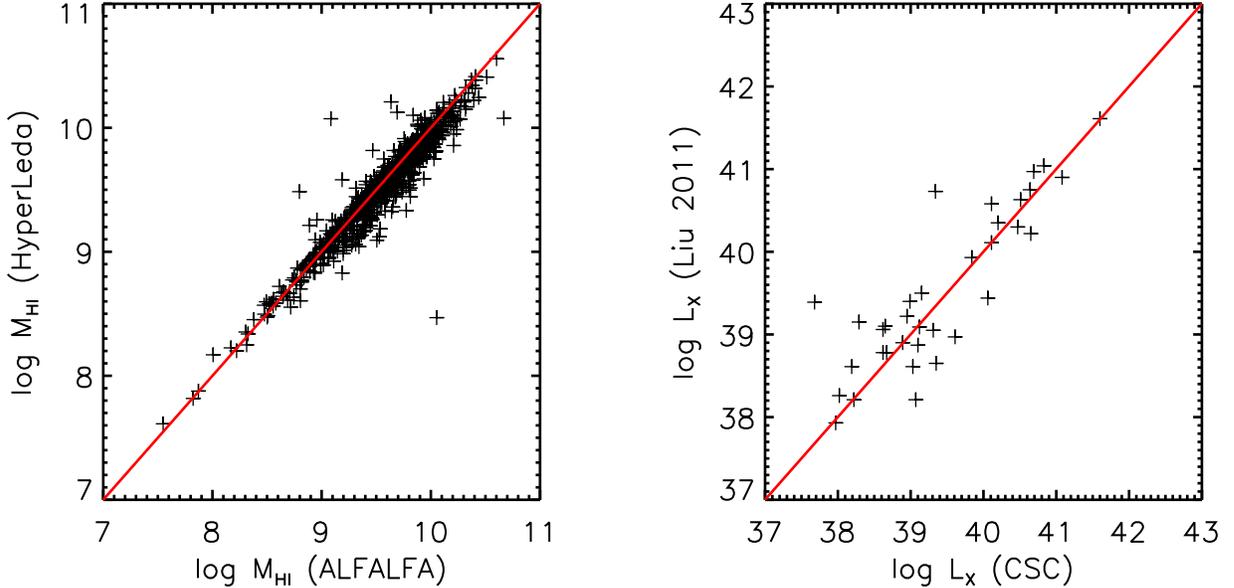} \figcaption{\small Left: HI
  masses from Hyperleda against new ALFALFA HI measurements. Right:
  X-ray luminosity from Liu (2011) against values from the Chandra
  Source Catalog. The red line indicates equal measurement values.}
\end{figure*}

We calculate the star formation rate (SFR) for the final sample of 129
spirals by cross-matching to the All-Sky GALEX UV
catalog\footnote{{\tt
    http://archive.stsci.edu/prepds/gcat/gcat\_casjobs.html}}. A match
radius of 6$''$ produces 46 galaxies with both FUV and NUV
magnitudes. UV fluxes trace star formation rate because they include
contributions from high-mass short-lived hot stars, and UV-derived
SFRs are not systematically offset from H$\alpha$ estimates (Bell \&
Kennicutt 2001).\footnote{While FUV and H$\alpha$ derived SFRs agree
  when internal extinction is taken into account for normal spirals
  such as considered here, this is not necessarily the case for dwarfs
  (Lee et al.~2009; McQuinn et al.~2015).} We use equations 6, 7, and
8 from Salim et al.~(2007). The FUV magnitude is corrected for
internal extinction using the FUV-NUV color as

\begin{equation*}
FUV_{\rm corr} = FUV - 2.99\times(FUV-NUV) - 0.27
\end{equation*}

and also corrected for a typical Milky Way reddening of 0.2
magnitudes.\footnote{While the SFRs are not strongly dependent on the
  sightline through the Milky Way, for a check of seven random
  galaxies the median, minimum, maximum, and standard deviation
  Galactic reddening is 0.24, 0.10, 0.44, and 0.12. These are
  calculated from Landolt $A_{\rm V}$ given in NED, taking $R_{\rm
    V}=3.1$ to find $E(B-V)$, then using the Fitzpatrick (1999)
  parameterization, implemented in IDL as {\tt fm\_unred}, at the FUV
  effective wavelength of 1516\AA.} After converting corrected FUV
magnitudes to flux densities and then luminosity densities (in units
of erg~s$^{-1}$~Hz$^{-1}$), the SFR in units of $M_{\odot}$~yr$^{-1}$,
is calculated as

\begin{equation*}
\log{SFR} = \log{L_{\rm FUV,corr}}  - 28.165
\end{equation*}

A similar recipe for calculating SFRs is given by Lee et al.~(2011;
see also Karachentsev \& Kaisina 2013) as $\log{SFR} = 2.78 -
0.4f_{\rm uv} + 2\log{d}$ with $f_{\rm uv}$ the extinction-corrected
FUV magnitude and $d$ the distance in Mpc. The All-Sky GALEX UV
catalog magnitudes are slightly fainter than the elliptical aperture
photometry done by Lee et al.~(2011); this is corrected by increasing
our SFRs by 0.03 dex, which is the mean difference for overlapping
galaxies at $d<10$ Mpc.

We verified that the late-type classifications from HyperLeda were
correct through inspection of SDSS cutouts (Figure~1).

\subsection{HI masses}

The ALFALFA survey has a limiting two-pass sensitivity of 1.8 mJy per
beam and a half power beam width of 3.3$'$x3.8$'$ for each of seven
feeds \citep{2005AJ....130.2598G}. Nearly all ALFALFA HI detected
galaxies have an optical counterpart (Haynes et al.~2011).  The
ALFALFA catalog uses source codes to classify objects. Code 1 objects
have a high signal-to-noise ratio ($S/N>6.5$) and are high confidence
detections. Objects with a source code of 2 have a moderate
signal-to-noise ratio ($4<S/N<6.5$) with an optical counterpart at the
same redshift, while entries with a source code of 3 also have a
moderate signal-to-noise ratio but without an optical counterpart;
deeper follow-up radio observations indicate that code 2 objects are
generally reliable while code 3 events are generally spurious. Code 4
and 5 objects have lower signal to noise ratios or RFI contamination
and are not reliable sources. Finally, code 9 objects are likely high
velocity clouds within the Milky Way. We used ALFALFA sources with
code 1 and 2 only.

To get HI data, we matched the resulting HyperLeda source positions to
those in $\alpha$70 with a search radius of 1 arc-min. Even with the
ALFALFA beam size ($\sim$50 projected kpc for a source at 50 Mpc)
source confusion is minimal, and the mass of the spiral dominates over
any small satellite galaxies. We retrieved HI fluxes from both
$\alpha$70 and HyperLeda literature when possible. We examined all
galaxies labeled as multiple in HyperLeda, and crudely categorized
their degree of interaction on a scale from 0--3, where 3 indicates
merging. (We also eliminated NGC 3314A at this stage because it is
superimposed on another galaxy.) Of the 129 galaxies in the final
sample, only NGC 3256 and NGC 7727 seem to be in the process of
merging. There are an additional 17 galaxies that are multiple, of
which 5 (NGC2993, NGC3395, NGC5194, NGC5427, and NGC5954) show clear
interaction signatures such as tidal tails.

The HI mass was calculated from the relationship

\begin{equation*}
M_{\rm HI} = 2.36\times10^{5}d_{\rm Mpc}^{2}S_{\rm HI}
\end{equation*}

and is expressed as a logarithm. From the parent sample there are 3966
spirals that have either ALFALFA or HyperLeda literature HI
measurements. In Figure 3 (left panel), we plot HI masses derived from
each catalog for 780 sources with data in both. For this comparison
for all sources we used redshift distances from HyperLeda. The
agreement between catalogs is excellent; the difference has median,
mean, and standard deviation of 0.003, $-0.008$, and 0.158 dex. A few
of the individual literature measurements in HyperLeda may have been
taken at higher angular resolution than the ALFALFA survey, and
resolve out flux; we prioritize ALFALFA data.

\subsection{X-ray luminosities and limits}

For X-ray coverage, {\it Chandra\/} is the optimal observatory with
which to determine nuclear X-ray luminosities for our sample of local
late-type galaxies due to its high angular resolution, which helps
avoid contamination from X-ray binaries that may be present throughout
the galaxy including near the (projected) center. Archival
observations from {\it ROSAT\/} or {\it XMM-Newton\/} do not provide
sufficient angular resolution and consequently are not considered. An
inherent limitation of this dataset is that most of the galaxies with
{\it Chandra\/} coverage were targeted rather than serendipitous and
we cannot control for the original motivation for targeting these
particular galaxies, or for the inhomogeneous sensitivities.

The {\it Chandra\/} Source Catalog \citep[CSC;][]{2010ApJS..189...37E}
is a list of X-ray detected point sources from archived Chandra
observations. The CSC contains many parameters including X-ray fluxes
for each source. Where an X-ray source is not present at the specified
coordinates, the catalog provides upper limit sensitivities, which we
include in our analysis. The CSC includes imaging spectroscopy (ACIS-I
or ACIS-S without gratings) data that is public as of 2011.

The HyperLeda source positions were matched to CSC X-ray sources with
a search radius of 2 arc-seconds to obtain only nuclear matches. The
X-ray luminosities were calculated from CSC full-band source fluxes
using HyperLeda distance moduli for each source as simply $L_{\rm X} =
4{\pi}d^{2}f_{\rm X}$ (i.e., no $k$-correction is required for these
low-redshift objects) and are also expressed as logarithms. For
sources that had {\it Chandra\/} coverage but were not detected in the
CSC, we retrieved the sensitivities at those positions. The
sensitivities were then used as upper limits in the fitting.

We also obtained nuclear X-ray luminosities from the catalog compiled
by Liu (2011). The agreement between the CSC and Liu luminosities is
also generally good (Figure 3, right), with offset
median/mean/standard deviation for 35 objects of $-0.11/-0.10/0.52$. 

\begin{figure*}
\includegraphics[scale=0.4]{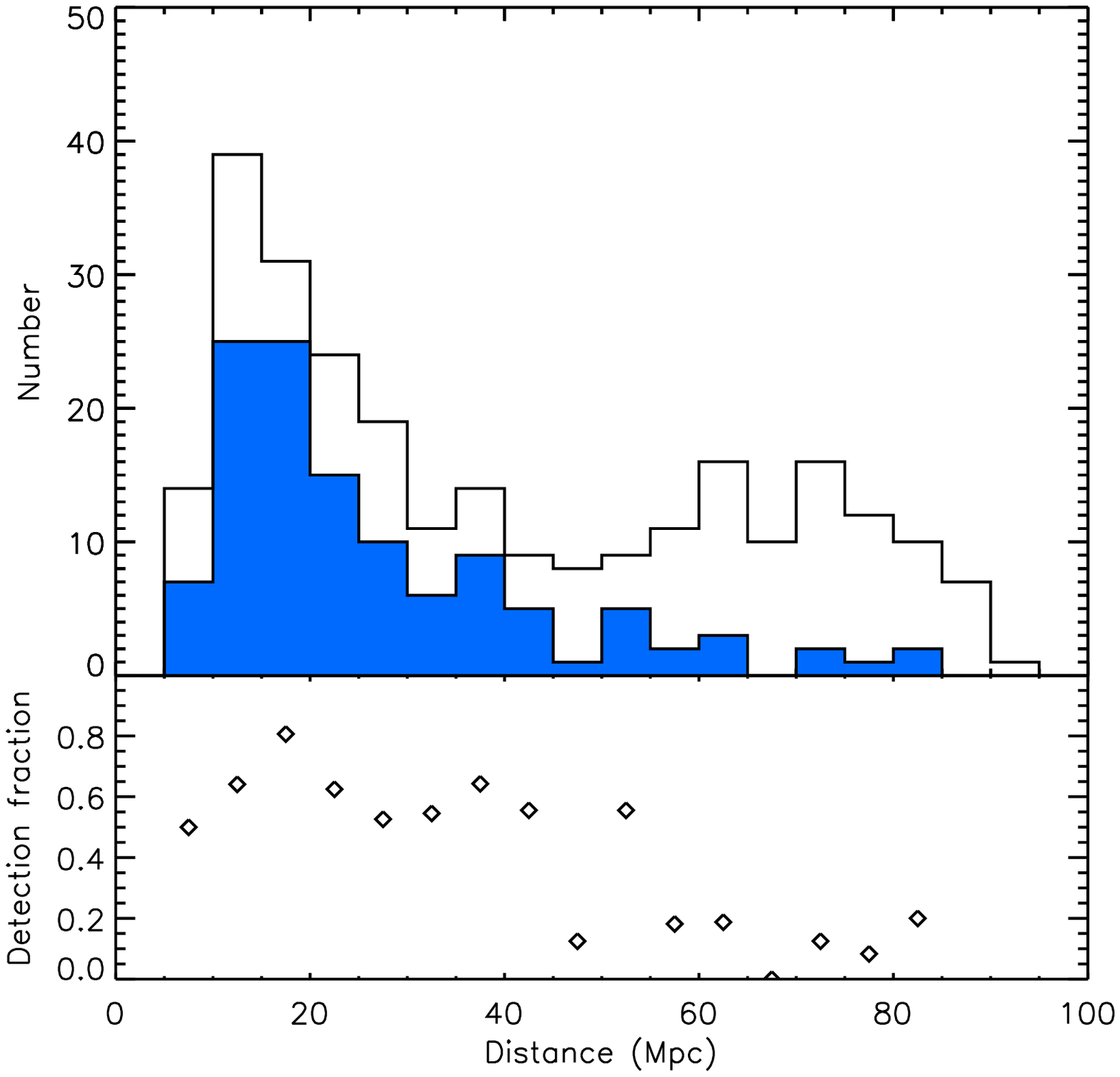}
\includegraphics[scale=0.58]{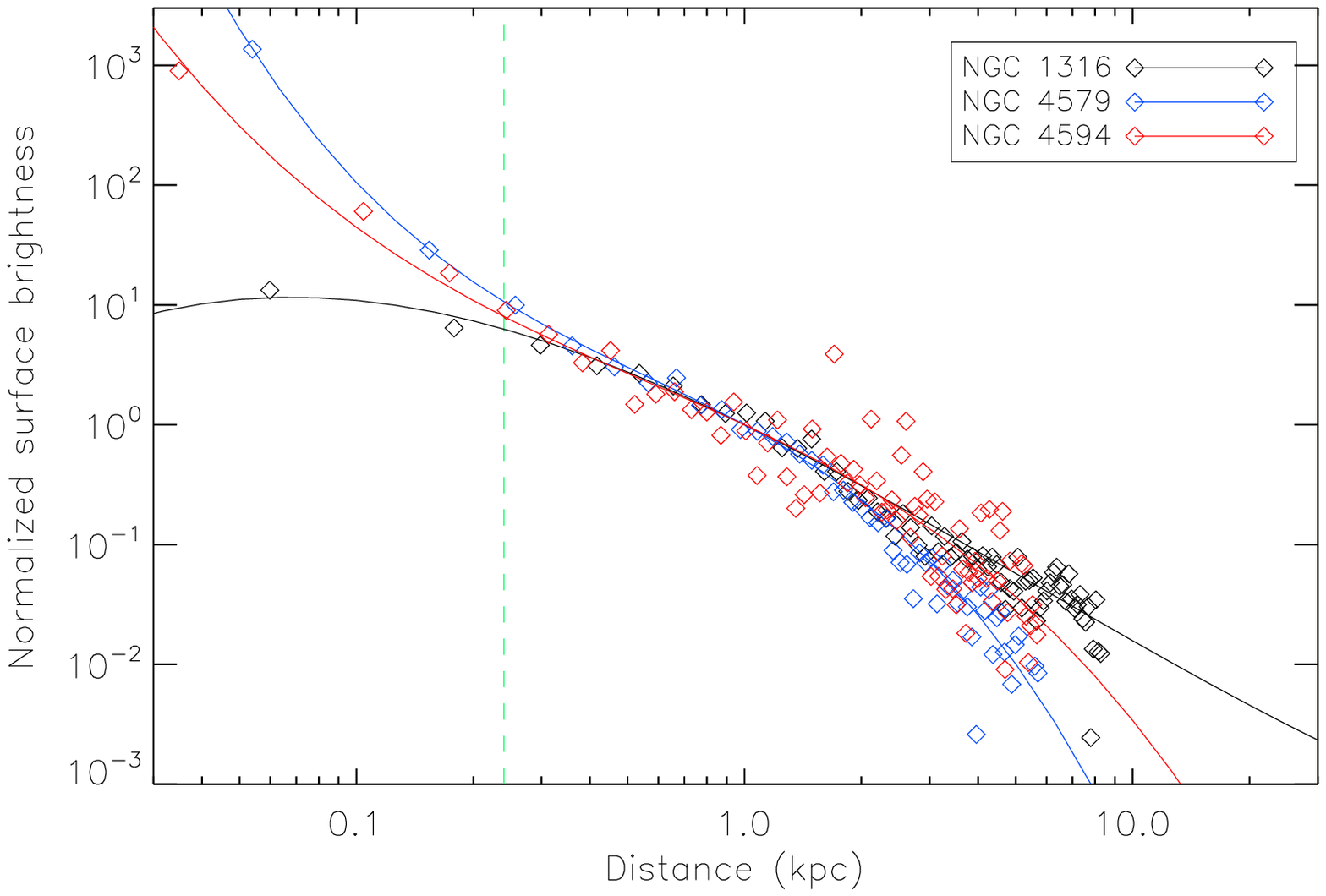} \figcaption{\small
  {\it Left:\/} Histogram of {\it Chandra\/} X-ray coverage (open) and
  nuclear X-ray source detection (filled) as a function of
  distance. We impose a cut of $d<50$ Mpc to maintain a high detection
  fraction. {\it Right:\/} X-ray surface brightness profiles for three
  example galaxies. NGC 1316 (Fornax A) is a complex quasi-lenticular
  with radio jets and a weak nuclear X-ray source, excluded from our
  sample. NGC 4579 is an Sb and NGC 4594 is an Sa spiral; both are
  also Seyfert 2 galaxies and contain a nuclear X-ray source powered
  by low-level supermassive black hole activity. The vertical line
  shows 1$''$ at 50 Mpc.}
\end{figure*}

For completeness, we also confirmed the X-ray flux measurements for a
random set of 15 sources with archival {\it Chandra\/} data. After
reducing the observations using the Chandra Interactive Analysis of
Observations (CIAO) software, we identified the central X-ray source
and extracted the flux from a circular aperture, then subtracted the
background. As expected, our hand-measured fluxes closely matched
those given in the catalog search. 

Our initial optical selection of the parent sample went out to a
distance of 90 Mpc. We examined the detection fraction of X-ray
nuclear sources as a function of distance (Figure 4, left) to see how
deep we could reasonably make our final sample. A detection refers to
a spiral in this Hyperleda sample that also has a Chandra CSC source
within 2$''$ of the optical position. It is apparent that there is a
significant decline in the detection fraction beyond 50 Mpc. We also
examined X-ray surface brightness profiles from a few deep {\it
  Chandra\/} observations of relatively nearby galaxies (Figure 4,
right) to map typical distributions of X-ray binaries (and hot gas) as
a function of radial distance. These profiles suggested that it is
possible to identify nuclear sources above the radial X-ray binary
contamination out to about 50 Mpc. (This does not guarantee that a
central source is associated with the SMBH, but the likelihood
increases with $L_{\rm X}>40$, which many of our galaxies surpass; see
Section 3.2) We therefore imposed $d<50$ Mpc for further analysis of
X-ray properties.

We rechecked the luminosity cut with the updated EDD distances, and
discarded UGC 05707, NGC 5474, and NGC 5585 as they no longer satisfy
$M_{B}<-18$. After examining Digitized Sky Survey images, we
additionally discarded NGC 4594, NGC 0672, and PGC 052935 as having
likely incorrect inclinations in HyperLeda (i.e., they are too edge-on
for our purposes). The resulting optical sample contains spiral
galaxies with B-band optical magnitudes $<-18$ within 50 Mpc that are
not edge-on.

Our final sample of spirals with {\it Chandra\/} coverage and HI data
then consists of 129 galaxies, of which 75 (58\%) have a central
source detected in X-rays. Table 1 lists the optical properties, HI
mass, and nuclear X-ray luminosity or limit for each galaxy.

\begin{figure*}
\includegraphics[scale=0.9]{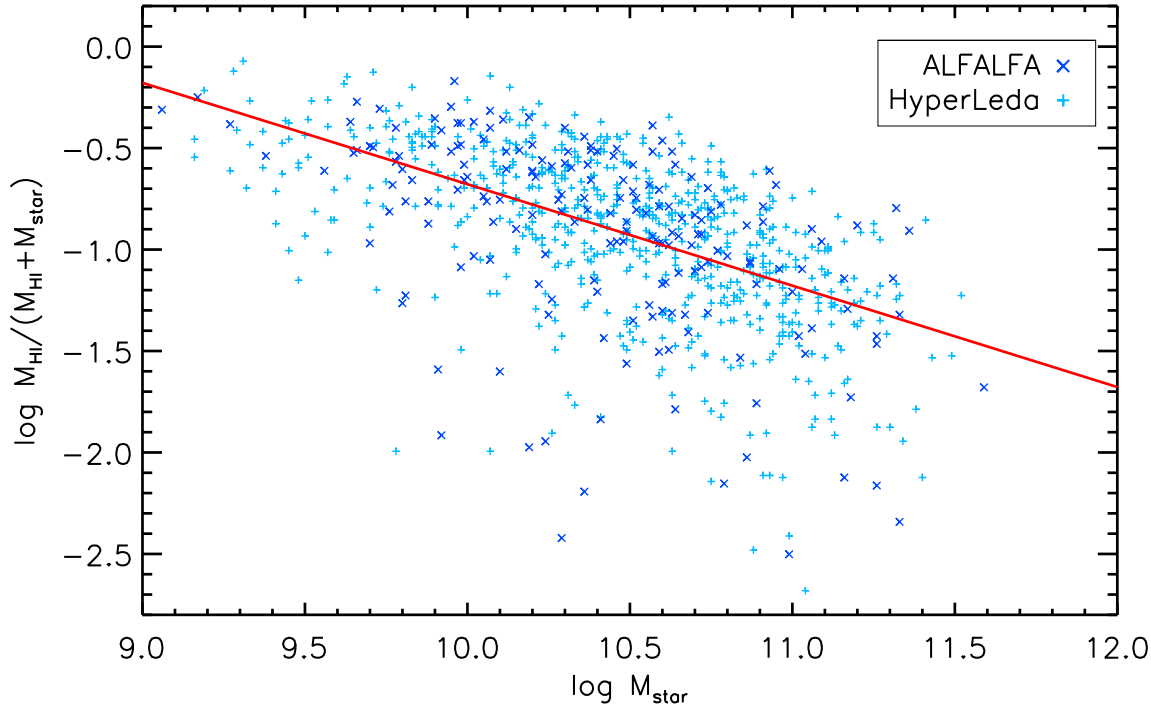} \figcaption{\small
  Comparison of HI to stellar mass, with fractional $M_{\rm
    HI}/(M_{\rm HI}+M_{\rm star})$ as a function of $M_{\rm
    star}$. There is a decline in the gas fraction of late-type
  galaxies with increasing stellar mass (e.g., Huang et al.~2012;
  Maddox et al.~2015). The solid red line shows the robust best-fit
  linear regression model.}
\end{figure*}

\section{Correlation analysis} 

We illustrate the relationship between stellar mass and cold gas
fraction (of the baryonic mass), here calculated as $M_{\rm
  HI}/(M_{\rm HI}+M_{\rm star})$, in Figure 5. We perform linear
regression using the IDL routine {\tt robust\_linefit.pro}, which uses
iterative bisquare weighting. The best-fit slope, calculated robustly
against outliers, is $-0.50\pm0.02$. Our data therefore show the
expected trend that optically brighter and more massive spirals
possess more HI gas in absolute terms (because the slope is $>-1$) but
have lower gas fractions (because the slope is $<0$). The correlation
is extremely significant with $p<0.001$. This trend has been
quantified using $\alpha$40 ALFALFA data by Huang et al.~(2012, 2014;
see also Maddox et al.~2015) and we here illustrate it with the new
$\alpha$70 catalog for nearby spirals. The slope of $\simeq-0.50$ that
we find for these galaxies is consistent with the slope of $-0.57$ for
$M_{\rm gas}/M_{\rm star}$ found by Peebles \& Shankar (2011); our
slope is slightly shallower because we examine gas mass in relation to
baryonic mass rather than only stellar mass.

There appear to be some outliers at low gas fractions in Figure 5. We
investigated the nature of galaxies that lie 2$\sigma$ above or below
the mean. Three galaxies are classified as outliers: NGC 4448 is red,
likely a star-forming spiral, NGC 4450 is a cluster galaxy in Virgo
but otherwise normal, and NGC 7727 is a late-stage merger with
tails. Each of these characteristics could be responsible for lowering
the galaxies’ gas fractions. It is interesting that there are no
outliers with high gas fractions, while lower than average gas
fractions are common for any given $M_{\rm star}$. It appears that gas
is easier to use than to store in spiral galaxies.

We also investigated the relationship between $M_{\rm HI}$ and $M_{\rm
  B}$. First we fit $M_{\rm HI}$ versus $M_{\rm B}$, then we fit
$M_{\rm B}$ versus $M_{\rm HI}$. Here we calculated the best-fit trend
using the bisector slope since it is not clear which direction the
causal relationship, if any, should go (Isobe et al.~1990). We find a
strong correlation, but this is not surprising since $M_{\rm B}$ is
used to calculate $M_{\rm star}$, so this is another perspective on
the gas fraction trend noted above.

\subsection{X-ray fitting}

We investigate nuclear X-ray luminosity versus stellar mass and HI
mass. We use the Bayesian linear regression code of Kelly (2007),
implemented in IDL as {\tt linmix\_err.pro}, to determine the
correlation slope and intercept taking both measurement uncertainties
and the X-ray upper limits into account. We report the preferred model
parameters as the median of 10000 draws from the posterior
distribution, with uncertainties corresponding to $1\sigma$ from the
16th and 84th percentiles.

For the stellar mass versus nuclear X-ray luminosity, fit as

\begin{equation*} 
\log{L_{\rm X}} - 39 = \alpha + \beta\times{(\log{M_{\rm star}} - 10.5)}
\end{equation*}

the best-fit intercept, slope, and intrinsic scatter are
$\alpha=0.18\pm0.13$, $\beta=1.73\pm0.26$, and $\sigma_{\rm
  0}=1.09\pm0.11$. While the intrinsic scatter is large, there is a
highly significant correlation between the variables, in the sense
that optically-brighter and more massive spiral galaxies also have
greater $L_{\rm X}$ values. The slope of $\beta=1.73\pm0.26$ is
consistent with $L_{\rm X}{\propto}M_{\rm star}^{2}$, which might be
expected for Bondi accretion where $\dot{m}{\propto}m^{2}$, if the
local mass $m$ were to scale with the total stellar mass $M_{\rm
  star}$, and further presuming that the black hole mass $M_{\rm BH}$
were a fixed fraction of $M_{\rm star}$ and that $L_{\rm X}$ scales
with $\dot{m}$. A slope of $\beta\sim1$, corresponding to a uniform
Eddington ratio under these assumptions, seems ruled out for our
spirals. Since these assumptions may not hold -- in particular, the
local gas supply may not be closely linked to the overall galaxy gas
content -- we refrain from drawing particular conclusions from the
best-fit slope.

For the HI mass versus nuclear X-ray luminosity, fit as

\begin{equation*} 
\log{L_{\rm X}} - 39 = \alpha + \beta\times{(\log{M_{\rm HI}} - 9.5)}
\end{equation*}

the best-fit intercept, slope, and intrinsic scatter are
$\alpha=-0.16\pm0.15$, $\beta=0.90\pm0.32$, and $\sigma_{\rm
  0}=1.33\pm0.13$. The intrinsic scatter is quite large, but there is
tentative evidence for a positive correlation between HI mass and
nuclear X-ray luminosity; the slope is greater than zero at the
$\sim$2.8$\sigma$ level. These correlations are shown in Figure 6 (top
panels) with the best-fit linear relations considering all sources
(solid lines). For completeness, we also show the regression results
for the galaxies with nuclear X-ray detections only (dashed
lines). {\it Chandra\/} detections and upper limits are indicated with
plus symbols and arrows, respectively.

\begin{figure*}
\includegraphics[scale=0.83]{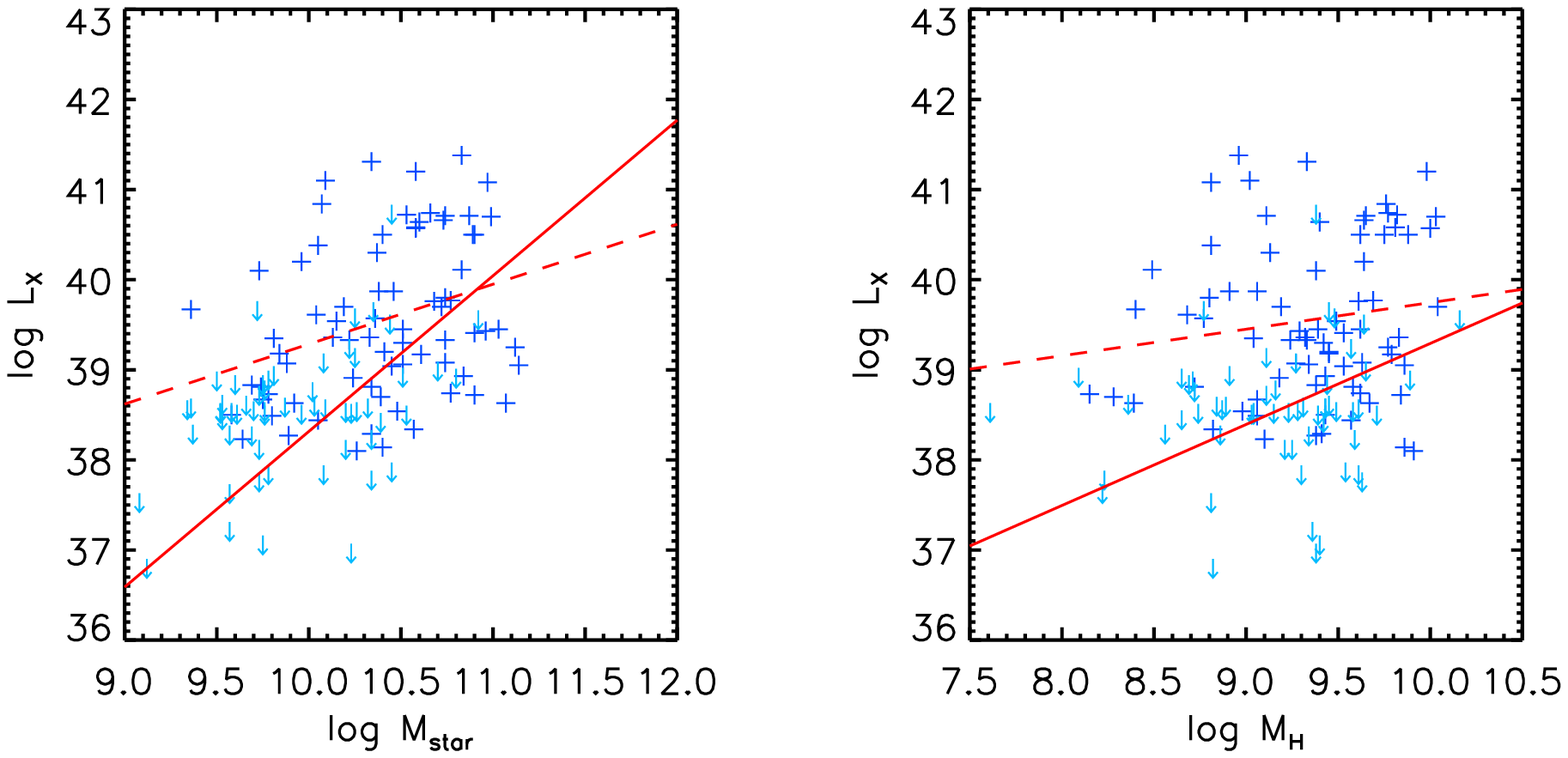}
\includegraphics[scale=0.83]{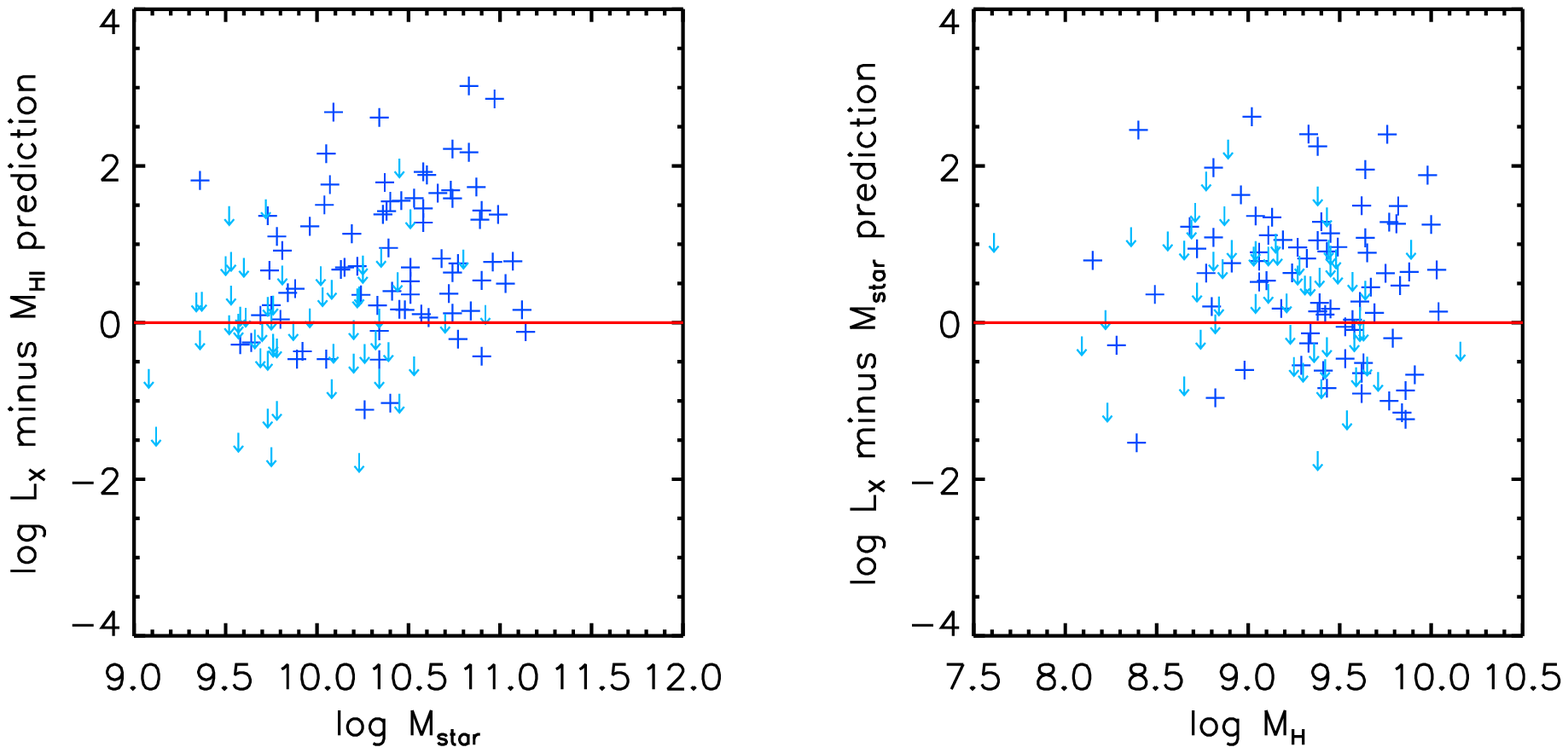} \figcaption{\small Top:
  X-ray luminosity versus stellar mass (left) and versus HI mass
  (right). The dashed lines are fits to detections only (plus
  symbols), while the solid lines are fits including the X-ray upper
  limits (down arrows). Measurement uncertainties of 0.2 dex and 0.1
  dex were assumed for $L_{\rm X}$ and both $M_{\rm star}$ and $M_{\rm
    HI}$, respectively. Linear regression was performed using the
  Bayesian methodology of Kelly (2007). Bottom: Residuals after
  subtracting the best-fit relations from the top panels. For
  reference, zero is marked with a solid horizontal red line.}
\end{figure*}

However, we then fit $L_{\rm X}$ as a function of both $M_{\rm star}$
and $M_{\rm HI}$ and found that the dependence upon $M_{\rm HI}$ in
this joint fit has a slope of $0.14\pm0.29$, consistent with zero. The
dependence upon $M_{\rm star}$ in the joint fit has a slope of
$1.69\pm0.30$. Further, the residuals of $L_{\rm X}-L_{\rm X}(M_{\rm
  star})$ show no trend with $M_{\rm HI}$. This contrasts with the
residuals for $L_{\rm X}-L_{\rm X}(M_{\rm HI})$ which still do show a
trend with $M_{\rm star}$ (Figure 6, bottom panels). The tentative
relationship between $L_{\rm X}$ and $M_{\rm HI}$ appears to be
entirely driven by $M_{\rm star}$. 

As a check, we also fit $L_{\rm X}$ as a function of both $M_{\rm
  star}$ and $M_{\rm HI}$ restricted to the 124 galaxies with
distances less than 40 Mpc. This subsample is in principle more
complete because it can include objects selected with redshift-based
distances $40<d<50$ Mpc but non-redshift distances $<40$ Mpc. The
slopes for $M_{\rm HI}$ and $M_{\rm star}$ are $0.16\pm0.29$ and
$1.60\pm0.29$, respectively, similar to fitting the full final
sample. Directly fitting $L_{\rm X}$ as a function of both distance
and $M_{\rm star}$ gives slopes of $0.02\pm0.01$ and $1.62\pm0.28$;
there is no significant direct dependence upon distance.

Finally, we used non-parametric methods to test for correlations as a
complementary check to our regression analysis. Kendall's tau values,
computed within ASURV, confirm the $L_{\rm X}(M_{\rm star})$
correlation at $p<0.0001$ and the $L_{\rm X}(M_{\rm HI})$ correlation
at $p<0.01$. We also used the Kendall's partial tau test developed by
Akritas and Siebert to test for the influence of the third variable on
the $L_{\rm X}$ correlations, finding that zero partial correlation is
rejected for $L_{\rm X}(M_{\rm star})$ but not for $L_{\rm X}(M_{\rm
  HI})$, using $p=0.05$ here as the dividing level. These results are
all consistent with the previous analysis and discussion.

\subsection{X-ray binary contamination}

An important caveat is that it is difficult to eliminate the chance of
high-mass X-ray binary contamination in late-type galaxies, even with
the outstanding angular resolution of {\it Chandra\/} (but see, e.g.,
Jenkins et al.~2011). The X-ray surface brightness profiles, and the
independent classification of Liu (2011), suggest that at least our
most luminous sources are indeed dominated by emission produced by the
SMBH. Here, we estimate potential X-ray binary contamination within
the {\it Chandra\/} aperture for each nuclear detection. 

Low-mass X-ray binaries (LMXBs) have a luminosity function that scales
with stellar mass (Gilfanov 2004), whereas high-mass X-ray binaries
(HMXBs) have a luminosity function that scales with star formation
rate (Grimm et al.~2003). This is because LMXBs are fed by Roche-lobe
overflow from their low-mass long-lived companion star whereas HMXBs
also accrete from the winds of their high-mass short-lived companion
star. A simplified empirical recipe is given by Lehmer et al.~(2008,
2010), who parameterize the total X-ray luminosity in binaries as a
function of both $M_{\rm star}$ and SFR. They find

\begin{equation*}
L_{\rm X,XRB} = 9.05\times10^{28}M_{\rm star} + 1.62\times10^{39}SFR,
\end{equation*}

with $M_{\rm star}$ and $SFR$ in units of $M_{\odot}$ and
$M_{\odot}/yr$, respectively. Note that this is for the hard 2--10 keV
band; to convert to the full 0.5--7 keV band over which the CSC fluxes
for our sources are measured, we add 0.11 dex to $L_{\rm X,XRB}$
(determined using PIMMS for a power-law spectrum with $\Gamma$=1.7 for
a typical column density of $N_{\rm H,gal}=2\times10^{20}$~cm$^{-2}$).

\begin{figure}
\includegraphics[scale=0.65]{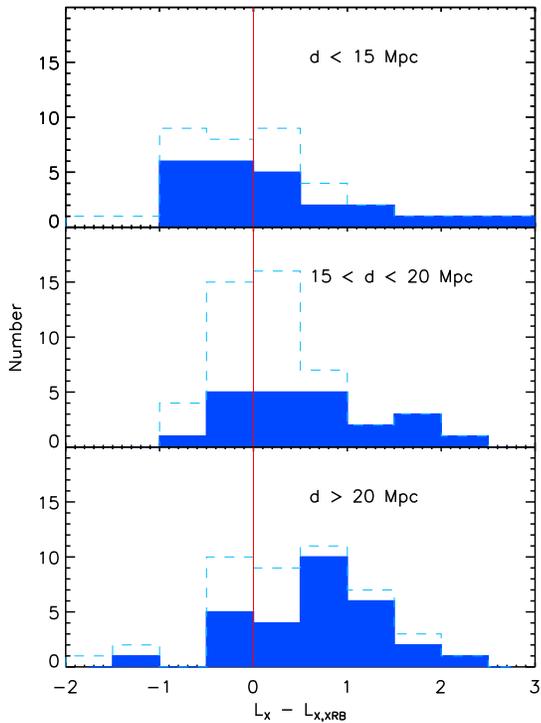} \figcaption{\small
  Difference between measured nuclear X-ray luminosities $L_{\rm X}$
  and the estimated X-ray binary contamination $L_{\rm X,XRB}$
  (including both from LMXBs and HMXBs) for detected sources (filled
  dark blue histogram) and including limits (dashed light blue), shown
  for three different bins of distance. Approximately 51/75 nuclear
  detections are likely associated with SMBH activity.}
\end{figure}

We conservatively calculate $L_{\rm X,XRB}$ within a 2$''$ aperture
centered on the galaxy, since the matching radius for nuclear {\it
  Chandra\/} sources is 2$''$. We assume that the star formation
throughout the galaxy follows the total light; this is also
conservative as these late-type galaxies tend to have bulges redder
than the disks. Rather than modeling the surface brightness profile
for each individual galaxy, we use a template to determine what
fraction of the total $M_{\rm star}$ and SFR (and hence what fraction
of the $L_{\rm X,XRB}$) is contained within 2$''$ as a function of
distance. The galaxy we select as a template is NGC 3344 which has
$\log{M_{\rm star}}=10.1$, $\log{SFR}=0.04$, $M_{\rm B}=-19.7$ (all
near the median values of our final sample), and is nearby at
$d\simeq$10~Mpc. We extract the optical surface brightness within
circular radial apertures (avoiding foreground stars) and model the
disk plus bulge as separate Sersic profile components (Ciotti \&
Bertin 1999) with $n=1$ and $n=4$, respectively. The scaling factor
for $L_{\rm X,XRB}$ based on this template changes smoothly from 13\%
to 61\% at distances from 5 to 50 Mpc; it may be parameterized as
$0.06917 + 0.01331{\times}d - 0.00005{\times}d^{2}$ with d in Mpc. A
galaxy-by-galaxy approach would provide some increase in accuracy but
using a template to calculate the fractional scaling suffices for our
statistical investigation. A histogram of expected values of $L_{\rm
  X,XRB}$ within 2$''$ is provided in Figure 7, along with the
measured nuclear X-ray luminosities for our galaxies.

We then estimate the probability that the nuclear X-ray emission is
associated with a SMBH rather than LMXBs or HMXBs for all sources with
a nuclear X-ray source detected by {\it Chandra\/}. This is calculated
as the likelihood that a random variable from a Gaussian distribution
in $\log{L_{\rm X}}$ with a mean of the scaled $\log{L_{\rm X,XRB}}$
and a sigma of 0.3 dex (i.e., scatter of 2) is less than or equal to
the measured nuclear $\log{L_{\rm X}}$ for that galaxy. Those values
are given in Table~1. Because most of the nuclear X-ray luminosities
are greater than those typically reached by LMXBs or even HMXBs, and
because the {\it Chandra\/} PSF out to our adopted distance limit only
encloses the central region of each galaxy, most X-ray detections are
secure. In particular, 51/75 detections have $L_{\rm X}>L_{\rm X,XRB}$
and probability $>0.5$. Revisiting the correlations of $L_{\rm X}$
with $M_{\rm HI}$ only, $M_{\rm star}$ only, and both fit jointly, the
slopes when treating galaxies with probabilities $<0.5$ as limits
rather than detections are similar but with larger uncertainties:
$0.65\pm0.42$, $1.74\pm0.40$, and $-0.01\pm0.40$ versus $1.80\pm0.47$,
respectively.

To incorporate quantitatively the possibility of LMXB and HMXB
contamination in our correlation analysis, we fit the sample 20
different times and probabilistically vary whether each X-ray source
is treated as a detection or a limit. For example, a nuclear X-ray
source with $L_{\rm X}=39.5$ and a probability that this is due to the
SMBH rather than LMXBs or HMXBs of 0.7 is treated as a detection in
70\% of the fits and as a limit in the remaining 30\%. The results of
these 20 fits are shown in Figure 8 for the fit of $L_{\rm X}$ as a
joint function of both $M_{\rm HI}$ and $M_{\rm star}$. In all cases,
the slopes show a significant dependance on $M_{\rm Star}$ and not on
$M_{\rm HI}$, and the preferred values are also within the
uncertainties from the previous fit treating all detections as
uncontaminated. We conclude that our results are robust against
typical LMXB and HMXB binary contamination. 

\begin{figure}
\includegraphics[scale=0.5]{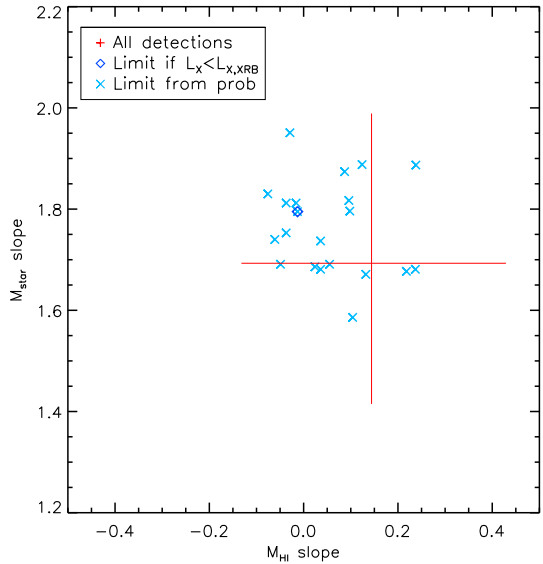} \figcaption{\small Slopes from
  the joint fit for $L_{\rm X}(M_{\rm HI}, M_{\rm star})$ for all
  detections (large red cross, showing uncertainties on the
  parameters) and then treating only galaxies with $L_{\rm X}>L_{\rm
    X,XRB}$ as detections (dark blue diamond), and then 20 random
  trials treating detections probabilistically (light blue x marks).}
\end{figure}

\subsection{Other potential dependencies}

\begin{figure*}
\includegraphics[scale=0.9]{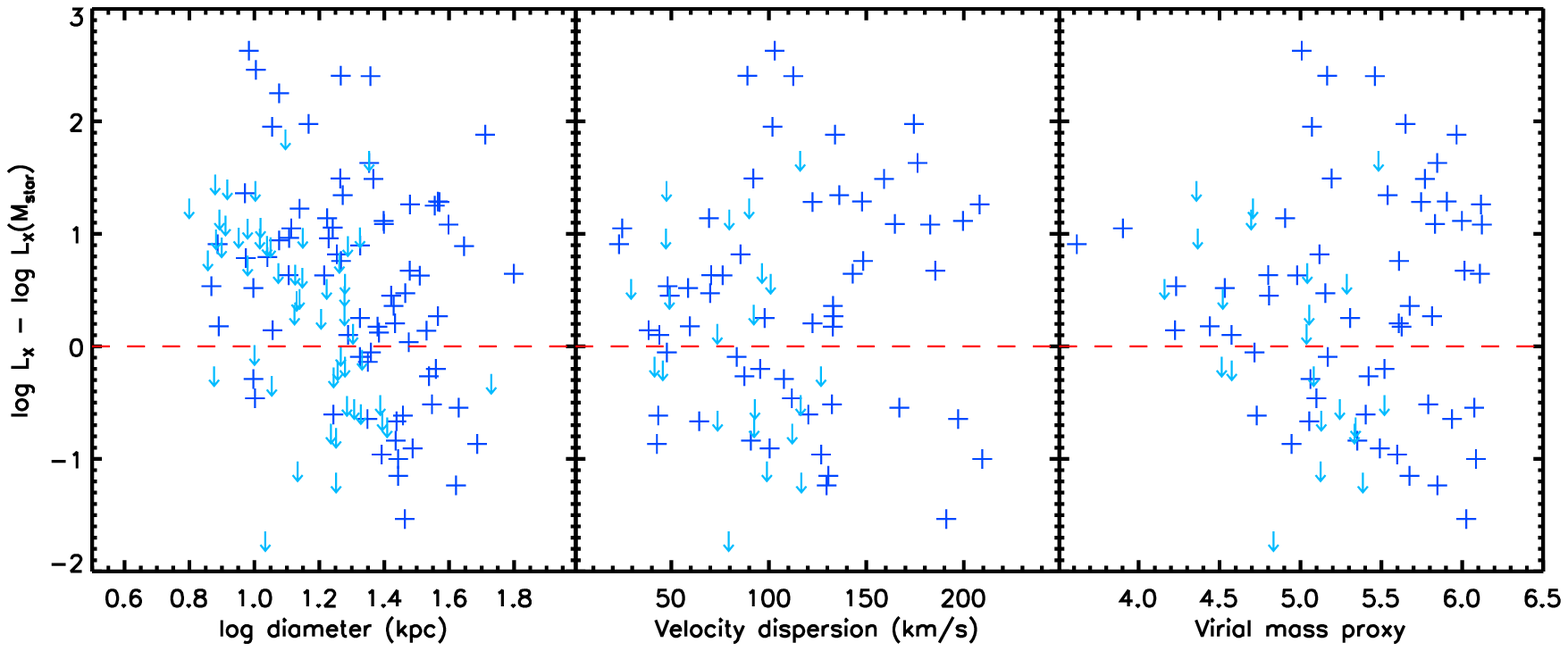}
\includegraphics[scale=0.9]{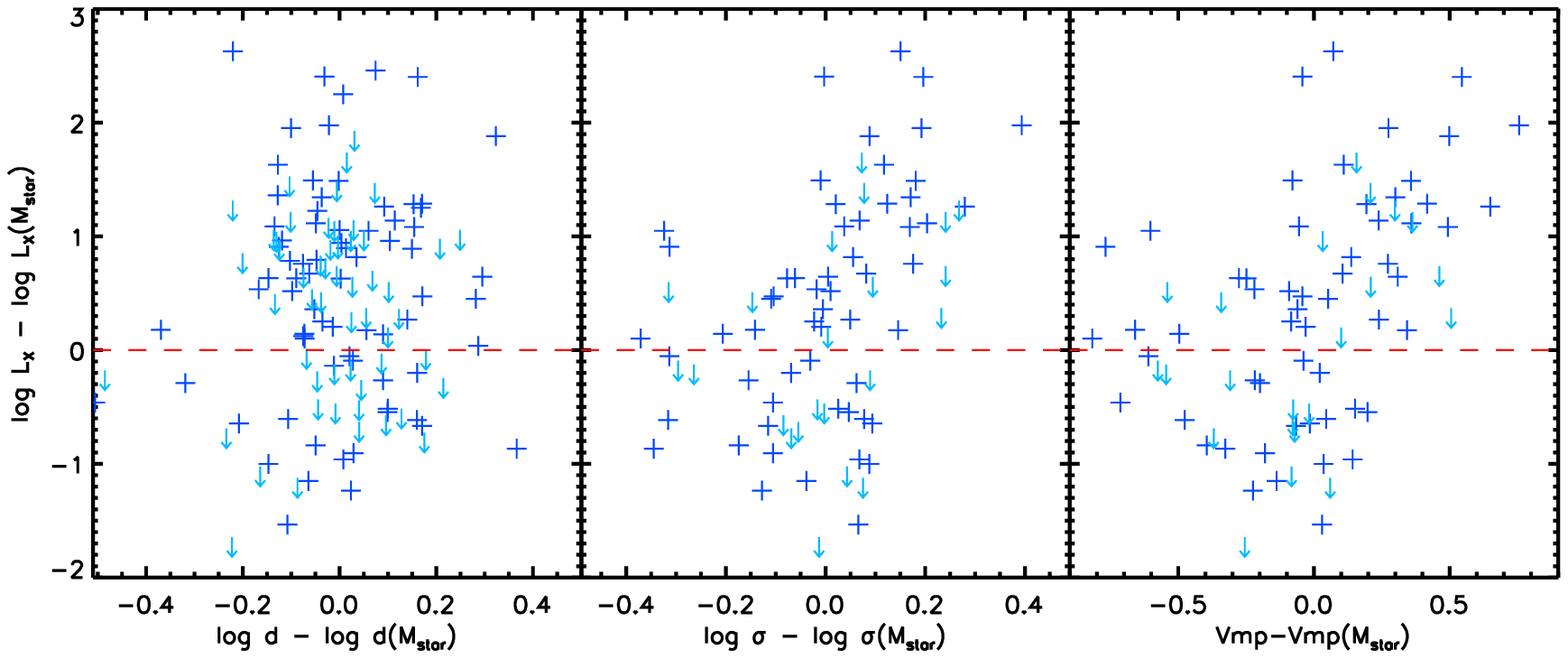} \figcaption{\small Top:
  Residual nuclear X-ray luminosity (after removing dependence on
  $M_{\rm star}$) versus galaxy size (left), central stellar velocity
  dispersion (center), and a virial mass proxy (right). For reference,
  the dashed red line shows zero residual X-ray luminsity. There are
  no significant correlations present. Bottom: As above, but here the
  galaxy size, central stellar velocity, and virial mass proxy are
  also corrected for their dependence on stellar mass.}
\end{figure*}

In Figure~9, top left panel, we show the relationship between residual
nuclear X-ray luminosity (i.e., after removing the dependence on
stellar mass, so $L_{\rm X}-L_{\rm X}(M_{\rm star})$) as a function of
galaxy size. Galaxy size is calculated from the apparent diameter
corrected for galactic extinction and inclination effects, as taken
from the HyperLeda database, and converted to physical size for the
distances given in Table~1. Presuming a uniform surface brightness
profile for our spiral galaxies (clearly incorrect but acceptable to
first order for the scaling arguments considered here), galaxy size is
a multiplicative constant times the effective radius. There appears to
be a possible tendency for more compact galaxies to be relatively more
X-ray luminous, but when we quantified this trend by fitting a linear
regression model using the {\tt linmix\_err.pro} IDL routine we found
that the slope was not significantly different from zero. It would be
interesting to revisit this point with a larger sample. We also tested
for a potential relationship between the central velocity dispersion
(again from HyperLeda) and residual nuclear X-ray luminosity, as shown
in the top central panel of Figure~9, and here too we do not find any
significant correlation. Finally, we consider the size multiplied by
the square of the velocity dispersion as a proxy for the virial mass,
as shown in the top right panel of Figure~9; here too we do not find
any significant correlation. We note that $L_{\rm X}$ might still
scale with any of these variables separately but if they in turn scale
with $M_{\rm star}$ then removing that dependence leaves no
trend. Finally, in the bottom panels of Figure~9, we again plot
residual nuclear X-ray luminosity against the galaxy size, central
stellar velocity, and virial mass proxy, but now correct these too for
their dependence on stellar mass. There is a possible tendency for
galaxies that have larger residual central stellar velocity
dispersions or virial masses, at a given stellar mass, to also have
relatively larger residual nuclear X-ray luminosities. SMBHs in local
galaxies correlate mostly with stellar velocity dispersion rather than
stellar mass (Shankar et al.~2016), and larger values of central
stellar velocity dispersion may indicate larger SMBHs (at a given
stellar mass) and hence greater X-ray luminosity at a given Eddington
ratio.

\section{Discussion and conclusions}   

\subsection{Eddington efficiencies}

We investigate the SMBH duty cycles for low and high $M_{\rm star}$
bins, through measuring the fraction of galaxies with $L_{\rm X}$
greater than 40 in a particular $M_{\rm star}$ range. We calculate the
fraction of galaxies with $L_{\rm X} > 40$ in the 9.3 – 10.3 $M_{\rm
  star}$ bin to be 5/67 = 7.5\% and the fraction of galaxies with
$L_{\rm X} > 40$ in the 10.3 – 11.3 $M_{\rm star}$ bin to be 18/62 =
29\%. The fraction of galaxies with $L_{\rm X} > 40$ within the entire
sample is 23/129 = 17.8\%. This is robust to potential XRB
contamination because the probabilities that the nuclear X-ray
emission is associated with a SMBH are $\simeq1$ for our galaxies with
$L_{\rm X}>40$.

For our sample of late-type galaxies, we use the M-sigma relationship
given in Gultekin et al.~(2009) to calculate black hole masses and
compare to $M_{\rm star}$: log($M_{\rm bh}/M_{\rm
  solar}$)=8.12+4.24$\log{\sigma}$/200 km/s).  We find on average
$M_{\rm bh} = M_{\rm star} – 3.22$ and use this to estimate black hole
mass for our sources without measured sigma. The bins of 9.3--10.3
$M_{\rm star}$ and 10.3--11.3 $M_{\rm star}$ then correspond to bins
of 6.1--7.1 $M_{\rm BH}$ and 7.1--8.1 $M_{\rm BH}$. For comparison,
Goulding et al.~(2010) use mid-IR selection techniques to calculate an
AGN fraction of about 20\% in their sample of nearby galaxies with
black hole masses similar to ours; see their Figure~5. This is in
rough agreement with our X-ray estimate of the fraction of SMBHs
accreting at $L_{\rm X} > 40$.

The late-type galaxies studied here have detected nuclear X-ray
luminosities ranging from $38.1 < \log{L_{\rm X}} < 41.4$. For a
typical SMBH of a few million solar masses, achieving $\log{L_{\rm
    X}}\sim40$ with a radiative efficiency of $\epsilon\sim0.1$ only
requires $\sim2\times10^{-6}$~M$_{\odot}$~yr$^{-1}$. These include
Seyfert AGNs as well as formally inactive galaxies; our sample probes
SMBHs across a continuum of low-level accretion rates.

\subsection{Multiwavelength activity indicators}

We were able to use indicators at a number of wavelengths to
distinguish SMBH activity from other potential sources of X-ray
emission. X-ray, optical and radio emissions can be used to probe
this. First, several galaxies are known Seyferts or LINERS or have
variable X-ray emission, which again improves the chances that we are
probing SMBH-linked activity rather than high-mass X-ray binary
contamination. Seyferts have strong optical features of AGN activity
that include strong emission lines as well as absorption lines from
the host galaxy. Seyferts 1 and 2 are thought to be the same objects
seen from different vantage points. Seyfert 1s have broadened emission
lines while Seyfert 2s do not because of our viewing angle. LINERS
have spectral features that could result from either low-level AGN
activity or star formation.

Table 1 shows optical indicators of black hole activity for selected
sources under the activity column. Several galaxies, including a high
fraction of the Seyferts, also have a central continuum 1.4 GHz point
source VLA detection (Ho \& Ulvestad 2001; also VLA FIRST survey data;
Becker et al.~1995). Radio detections serve as a signpost of SMBH
activity because, in the fundamental plane (e.g., Merloni et
al.~2003), the radio luminosity one expects is a function of the X-ray
luminosity and the SMBH mass. While X-ray binaries have radio
emission, at a given X-ray luminosity, a 10 solar mass XRB will have a
much weaker radio luminosity than a million solar mass SMBH. Star
formation processes and supernovae remnants can also produce radio
sources, albeit spread throughout the external galaxy rather than
centralized (note the typical angular resolution of these VLA maps is
$\sim$5$''$), but even compared to HMXBs supernova remnants have short
lifetimes of a few hundred thousand years when they are radio bright.

The average X-ray luminosities of the galaxies identified as Seyferts
or LINERS are larger than those not classified as such in NED. The
detection fraction for Seyferts and LINERS is 92\% (34/37) compared to
43\% for optically inactive galaxies, and 79\% of those detections
(27/34) have $L_{\rm X}>L_{\rm X,XRB}$. Taking X-ray limits into
account using ASURV (Isobe et al.~1990), Seyferts and LINERS have a
(Kaplan-Meier) mean $\log{L_{\rm X}}=39.83\pm0.16$ compared to
optically inactive galaxies at $\log{L_{\rm X}}=38.32\pm0.13$, and
two-sample tests indicate probabilities of $p<0.001$ that our samples
of Seyferts and LINERS versus optically inactive galaxies have the
same underlying $L_{\rm X}$ distribution. Similar results hold when
normalizing $L_{\rm X}$ by $M_{\rm star}$. This is consistent with the
optically inactive galaxies hosting SMBHs accreting at a lower rate
than in the Seyferts and LINERS. 

At the same time, the gas fractions of these Seyfert and LINER
galaxies are similar to (if anything perhaps slightly lower than)
those of optically inactive galaxies at matched $M_{\rm star}$
values. We checked this by drawing a random subsample without
consideration of gas fraction from the optically inactive galaxies
weighted to match the $M_{\rm star}$ distribution of the Seyferts and
LINERS (achieving good agreement in $M_{\rm star}$ after weighting; KS
test probability $p=0.5$). The resulting $M_{\rm star}$ matched
subsample of non-Seyferts has a distribution of gas fractions that is
consistent with the Seyferts (KS test $p=0.3$). This also suggests
that SMBH accretion rates are not dependent on galaxy-wide gas
fractions alone.

\subsection{SFR, cold gas, and SMBH activity}

Our fits test for a relationship between black hole accretion rate and
HI content. Since cold gas is available for star formation, we might
expect a positive trend based on the correlation between star
formation rate and AGN fraction found by Rafferty et al.~(2011). On
the other hand, Fabello et al.~(2011, 2012) find that bulge properties
and SMBH accretion rate are not linked to HI gas in blue galaxies
(although they do appear to correlate in red galaxies). Specifically,
the gas fraction in galaxies that host AGNs does not differ from that
in a matched comparison sample of galaxies, or depend on the
\ion{O}{3}/$M_{\rm BH}$ accretion indicator (Fabello et al.~2011). Our
results are consistent with this finding, to the extent that X-ray
luminosities provide a complementary (if not entirely overlapping)
view of nuclear activity. The lack of a strong dependence of $L_{\rm
  X}$ upon $M_{\rm HI}$ in our analysis suggests that the galaxy-wide
cold gas does not directly influence low-level supermassive black hole
activity in these spirals. 

We also investigated the relationship between the presence of bar
features (barred spirals) and the HI and X-ray luminosities. A bar
provides grativational torque and is believed to funnel gas down to
the central regions of the galaxy by decreasing its angular momentum
(e.g., Shlosman et al.~1990), so we might expect barred galaxies to
have greater nuclear $L_{\rm X}$ values at a given HI mass. We
identified barred galaxies using the classification in NED, as given
in Table~1. From our 132 late-type galaxies, only 8 lack bar
classifications in NED; 34 have no bar (type A), 39 are conclusively
barred (type B), and 51 have intermediate or weak bar (type AB).  We
do not find any significant difference in the average nuclear X-ray
luminosities of late-type galaxies with versus without
bars. Specifically, barred galaxies have a (Kaplan-Meier) mean
$\log{L_{\rm X}}=38.60\pm0.25$ compared to non-barred galaxies at
$\log{L_{\rm X}}=38.87\pm0.23$, and two-sample tests are consistent
($p>0.5$) with these samples having the same underlying $L_{\rm X}$
distribution. (Weakly barred galaxies have $\log{L_{\rm
    X}}=38.76\pm0.16$ and are also consistent with non-barred
galaxies.)  This suggests that bars may not channel a significant
amount of gas directly into the black hole or alternatively perhaps
that there is a balance between feeding and accretion. Our results are
consistent with the previous findings of Zhang et al.~(2009) that
barred galaxies do not have greater nuclear X-ray luminosities than
non-barred galaxies; if anything, those authors find that strongly
barred galaxies tend to be X-ray faint. This might also be related to
the results from Masters et al.~(2012) that barred galaxies are
generally less gas-rich, perhaps because the bar concentrates gas in
the galactic center where it is rapidly consumed in star formation.

We also investigate the potential difference in X-ray luminosity
between interacting and non-interacting galaxies. The nuclear X-ray
detection fraction is 13/19 (68\%) for galaxies flagged as
``multiple'' in HyperLeda, and 62/110 (56\%) for the rest. The average
X-ray luminosity for detected interacting/merging galaxies is
$39.63\pm0.28$ and average X-ray luminosity for the non-interacting
detected galaxies is $39.54\pm0.08$. Perhaps because we do not have a
large number of interacting galaxies in our sample, the difference in
detection fractions and average X-ray luminosity is suggestive but not
formally significant. Koss et al.~(2011) found that hard X-ray
detected AGN at $z<0.05$ exhibit a higher fraction of mergers.

\begin{figure}
\includegraphics[scale=0.5]{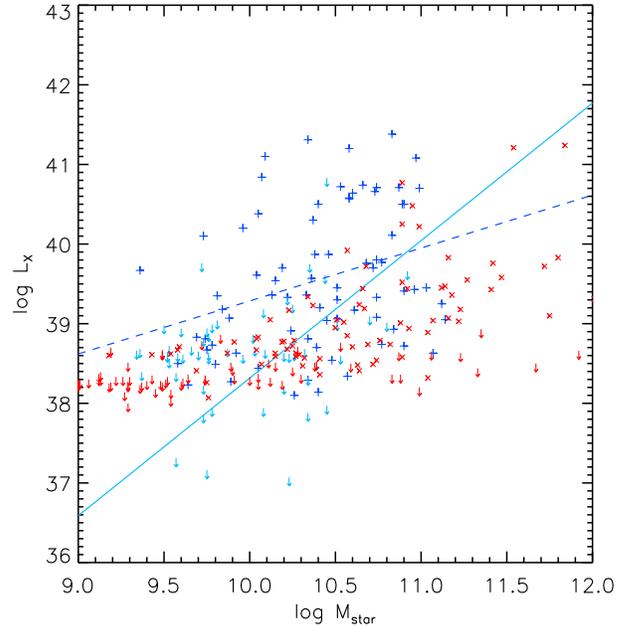} \figcaption{\small Nuclear
  X-ray luminosity as a function of galaxy stellar mass for our sample
  of late-type galaxies, along with the AMUSE sample of early-type
  galaxies. The early-type galaxies tend to have somewhat lower
  nuclear X-ray luminosities at a given stellar mass, and have a
  flatter slope when fit with a linear model.}
\end{figure}

It is interesting to compare these results to low-level SMBH activity
in early-type galaxies. Ellipticals also show a trend in nuclear
$L_{\rm X}$ with $M_{\rm star}$ (Pellegrini 2010; Gallo et al.~2010),
with an apparent enhancement in scaled $L_{\rm X}$ for galaxies in
sparser large-scale environments (Miller et al.~2012). A comparison of
our sample of late-type galaxies to the early-type galaxies in the
AMUSE project (Gallo et al.~2010; Miller et al.~2012) indicates that
the ellipticals tend toward lower values of $L_{\rm X}$ at a given
$M_{\rm star}$ (see Figure~10), and the correlation has a flatter
slope. Since galaxies with a particular $L_{\rm X}$ value can have a
wide range of stellar masses, it seems likely that there is a wide
range in SMBH accretion rates and/or efficiencies in these weakly
active systems. Because early-type galaxies in the field have more HI
than their cluster counterparts (Grossi et al.~2009; Oosterloo et
al.~2010), if still notably less in absolute terms than do late-type
galaxies, this suggests that cold gas does play a role in nuclear
activity. An analysis of dust features (from HST maps) or CO emission
(from IRAM) in early-type galaxies finds that the presence of dust or
CO is associated with higher Eddington ratios (i.e., $>10^{-7}$), but
that there is no correlation between the CO mass and the Eddington
ratio (Hodges-Kluck et al., in preparation). This is consistent with
our findings for late-type galaxies that the total amount of
galaxy-wide gas on kpc scales does not directly set the SMBH accretion
on pc scales, but supports the importance of cold gas as a
contributing element to SMBH growth. It also aligns with Sabater et
al.~(2015) finding that large-scale environment and galaxy mergers are
not directly responsible for AGN activity, but central cold gas supply
feeds the SMBH through secular processes, and with LaMassa et
al.~(2013) finding that central SFR correlates with AGN activity.

In summary, this work tests the hypothesis that more cold gas in a
galaxy leads to more accretion onto the central SMBH, on average; we
find that this is not supported, after accounting for the effect of
galactic stellar mass. This implies that the mechanisms that directly
supply SMBHs with gas near the galactic center do not know or care
about the total available gas throughout the galaxy. A lack of
correlation between galaxy-wide HI mass and nuclear X-ray emission
suggests that the central $\sim$50 pc are most relevant for SMBH
feeding and that the local gas supply is only loosely linked to the
outskirts. Future high angular resolution observations of the gas
content and dynamics in the imminent vicinity of the SMBH, for example
conducted with ALMA, could help establish a link between local cold
gas and SMBH activity.

\acknowledgments 

We gratefully acknowledge the UAT team for support of this work and
preliminary $\alpha$70 catalog. We also thank the NSF for partial
support through grant AST-1211005. We thank the referee for helpful
comments that improved this paper.

We acknowledge the usage of the HyperLeda database ({\tt
  http://leda.univ-lyon1.fr}). This research has made use of the
NASA/IPAC Extragalactic Database (NED) which is operated by the Jet
Propulsion Laboratory, California Institute of Technology, under
contract with the National Aeronautics and Space Administration. This
research has made use of data obtained from the Chandra Source
Catalog, provided by the Chandra X-ray Center (CXC) as part of the
Chandra Data Archive.


\LongTables
\begin{deluxetable*}{p{50pt}rrrrrrrrrrrrrrrr}
\tablecaption{Sample Properties}
\tabletypesize{\scriptsize}
\tablecolumns{9}
\tablewidth{17.5cm}

\tablehead{\colhead{Name} & \colhead{RA} & \colhead{Dec} & \colhead{Dist} & \colhead{$M_{\rm B}$}
  & \colhead{$M_{\rm HI}$} & \colhead{$M_{\rm star}$} & \colhead{$L_{\rm X}$} & \colhead{Prob} 
  & \colhead{Bar} & \colhead{Mult} & \colhead{Act} \\
\colhead{} & \colhead{(deg)} & \colhead{(deg)} & \colhead{(Mpc)} & \colhead{} & \colhead{($M_{\odot}$)}
  & \colhead{($M_{\odot}$)} & \colhead{(erg/s)} & \colhead{} & \colhead{} & \colhead{} & \colhead{} \\[-0.1cm]}

\startdata 

     NGC0255 &  11.947050 & -11.468730 & 18.0 & -19.1 &  9.4 &  9.7 & $<$38.7 &      & AB &    &    \\
     NGC0470 &  19.936800 &   3.409870 & 26.9 & -20.1 &  9.4 & 10.4 & $<$40.8 &      & A  & M1 & F  \\
     NGC0628 &  24.174000 &  15.783320 &  9.0 & -20.4 &  9.9 & 10.3 & 38.1    & 0.04 & A  &    &    \\
     NGC0685 &  26.928300 & -52.761790 & 16.0 & -19.4 &  9.4 &  9.8 & $<$38.6 &      & AB &    &    \\
    UGC01378 &  29.080050 &  73.282780 & 46.3 & -21.7 & 10.2 & 10.9 & $<$39.7 &      & B  &    &    \\
     NGC0925 &  36.820350 &  33.579000 &  9.2 & -20.0 &  9.6 & 10.1 & 38.4    & 0.49 & AB &    &    \\
     NGC0949 &  37.702950 &  37.136590 & 11.3 & -18.6 &  8.7 &  9.5 & $<$38.5 &      & A  &    &    \\
     NGC0988 &  38.866050 &  -9.356420 & 17.3 & -21.0 &  9.2 & 10.2 & $<$38.6 &      & B  &    &    \\
     NGC0991 &  38.885850 &  -7.155020 & 17.3 & -18.4 &  9.0 &  9.6 & $<$38.6 &      & AB &    &    \\
      IC0239 &  39.115950 &  38.969030 & 10.0 & -18.5 &  9.2 &  9.7 & $<$38.2 &      & AB &    &    \\
     NGC1055 &  40.437300 &   0.443180 & 12.3 & -19.6 &  9.6 & 10.3 & $<$38.3 &      & B  &    & F  \\
     NGC1068 &  40.669650 &  -0.013240 & 12.3 & -20.9 &  9.0 & 10.8 & 41.4    & 1.00 & A  &    & Sy, F\\
     NGC1058 &  40.875450 &  37.341100 &  9.9 & -18.7 &  9.1 &  9.6 & 38.2    & 0.26 & A  &    & Sy \\
     NGC1073 &  40.918800 &   1.375700 & 12.3 & -19.3 &  9.4 &  9.7 & 38.8    & 0.98 & B  &    &    \\
     NGC1097 &  41.579550 & -30.274970 & 14.2 & -21.1 &  9.7 & 10.9 & 40.7    & 1.00 & B  &    & L  \\
     NGC1341 &  51.993300 & -37.150060 & 19.0 & -18.7 &  8.4 &  9.5 & $<$38.7 &      & AB &    &    \\
     NGC1365 &  53.401650 & -36.140510 & 18.0 & -21.4 &  9.9 & 10.9 & 40.5    & 1.00 & B  &    & Sy \\
     NGC1367 &  53.755350 & -24.933230 & 22.7 & -20.5 &  9.8 & 10.9 & 40.5    & 1.00 & AB &    &    \\
     NGC1493 &  59.364450 & -46.210900 & 16.8 & -19.4 &  9.3 &  9.9 & 39.1    & 0.98 & B  &    &    \\
     NGC1637 &  70.367400 &  -2.858090 &  9.8 & -18.7 &  9.1 &  9.8 & 38.7    & 0.93 & AB &    &    \\
     NGC1640 &  70.560600 & -20.434740 & 16.8 & -18.9 &  8.7 & 10.0 & 39.6    & 1.00 & B  &    &    \\
     NGC1703 &  73.217250 & -59.742190 & 11.9 & -18.6 &  8.9 &  9.6 & $<$38.4 &      & B  &    &    \\
      IC0396 &  74.495700 &  68.323520 & 14.0 & -18.9 &  8.2 &  9.8 & 38.7    & 0.90 &    &    &    \\
     NGC2082 &  85.462950 & -64.301080 & 13.1 & -18.2 &  8.6 &  9.4 & $<$38.4 &      & B  &    &    \\
     NGC2139 &  90.283350 & -23.672410 & 21.0 & -20.0 &  9.6 &  9.8 & $<$37.9 &      & AB &    &    \\
     NGC2276 & 111.804450 &  85.751390 & 30.3 & -21.1 &  9.6 & 10.3 & 38.8    & 0.19 & AB & M1 &    \\
     NGC2500 & 120.471600 &  50.737230 & 15.0 & -18.9 &  9.1 &  9.7 & $<$38.8 &      & B  &    &    \\
     NGC2541 & 123.667200 &  49.061580 & 11.2 & -18.7 &  9.4 &  9.4 & $<$38.7 &      & A  &    &    \\
     NGC2782 & 138.521100 &  40.113770 & 42.1 & -21.0 &  9.6 & 10.7 & 40.7    & 1.00 & AB &    & Sy, F\\
     NGC2841 & 140.510250 &  50.976830 & 14.1 & -21.2 &  9.8 & 11.1 & 39.2    & 0.13 & A  &    & L, F\\
     NGC2906 & 143.026050 &   8.441590 & 33.6 & -20.1 &  9.1 & 10.2 & $<$39.2 &      &    &    &    \\
     NGC2993 & 146.451300 & -14.368380 & 29.0 & -19.6 &  9.4 &  9.7 & 40.1    & 1.00 &    & M2 &    \\
     NGC3066 & 150.545400 &  72.125250 & 28.0 & -19.0 &  9.0 &  9.8 & 39.4    & 0.99 & AB &    &    \\
     NGC3169 & 153.562050 &   3.466360 & 23.7 & -21.0 & 10.0 & 11.0 & 40.7    & 1.00 & A  & M1 & Sy, F\\
      IC2560 & 154.077750 & -33.563870 & 32.5 & -20.9 & 10.0 & 10.6 & 40.6    & 1.00 & B  &    & Sy \\
     NGC3185 & 154.410750 &  21.688320 & 26.4 & -19.5 &  8.8 & 10.4 & 39.6    & 0.96 & B  &    & Sy, F\\
     NGC3184 & 154.570800 &  41.424360 & 13.0 & -20.2 &  9.4 & 10.3 & 38.3    & 0.04 & AB &    &    \\
     NGC3256 & 156.963750 & -43.903760 & 35.2 & -21.4 &  9.8 & 10.7 & 40.7    & 1.00 &    & M3 &    \\
    NGC3314A & 159.303450 & -27.683760 & 50.6 & -20.0 &  9.5 & 10.2 & $<$39.7 &      &    &    &    \\
     NGC3310 & 159.689850 &  53.501740 & 20.0 & -20.4 &  9.6 & 10.0 & 40.2    & 1.00 & AB &    &    \\
     NGC3344 & 160.879650 &  24.922150 & 10.0 & -19.7 &  9.6 & 10.1 & $<$38.7 &      & AB &    & F  \\
     NGC3368 & 161.690400 &  11.819810 &  7.2 & -19.6 &  9.0 & 10.5 & 38.5    & 0.21 & AB &    & F  \\
     NGC3389 & 162.116400 &  12.533130 & 21.4 & -19.9 &  9.4 &  9.8 & 39.2    & 0.98 & A  &    &    \\
     NGC3395 & 162.458850 &  32.982830 & 25.7 & -20.2 &  9.5 &  9.8 & $<$38.6 &      & AB & M2 &    \\
     NGC3507 & 165.855750 &  18.135990 & 20.9 & -19.7 &  9.3 & 10.1 & 39.4    & 0.98 & B  &    & F  \\
     NGC3521 & 166.452450 &  -0.035780 & 11.2 & -21.0 &  9.8 & 10.9 & 38.7    & 0.03 & AB &    &    \\
     NGC3568 & 167.702400 & -37.447790 & 40.4 & -21.2 & 10.0 & 10.7 & 39.7    & 0.73 & B  &    &    \\
     NGC3556 & 167.878950 &  55.674150 &  9.6 & -19.9 &  9.4 & 10.2 & $<$37.1 &      & B  &    &    \\
     NGC3627 & 170.062500 &  12.990990 &  8.3 & -20.5 &  8.8 & 10.6 & 38.3    & 0.03 & AB & M0 & L, F\\
     NGC3631 & 170.262000 &  53.169940 & 18.0 & -20.8 &  9.4 & 10.5 & 39.3    & 0.78 & A  &    &    \\
     NGC3683 & 171.882600 &  56.876990 & 25.9 & -19.8 &  9.5 & 10.2 & 39.5    & 0.96 & B  &    &    \\
     NGC3718 & 173.145150 &  53.067870 & 17.1 & -20.0 &  9.8 & 10.5 & 40.7    & 1.00 & B  & M1 & L, F\\
     NGC3887 & 176.769150 & -16.854710 & 18.2 & -20.2 &  9.4 & 10.3 & $<$38.7 &      & B  &    &    \\
     NGC3898 & 177.313650 &  56.084030 & 22.1 & -20.8 &  9.6 & 11.0 & 39.5    & 0.27 & A  &    & F  \\
     NGC3913 & 177.662250 &  55.353900 & 17.1 & -18.0 &  8.9 &  9.3 & $<$38.7 &      & A  &    &    \\
     NGC3938 & 178.205700 &  44.120800 & 17.1 & -20.4 &  9.6 & 10.3 & $<$38.6 &      & A  &    &    \\
     NGC3982 & 179.117250 &  55.124930 & 20.5 & -19.9 &  9.2 & 10.2 & 39.3    & 0.95 & AB &    & Sy, F\\
    UGC06930 & 179.323050 &  49.283720 & 17.1 & -18.6 &  9.3 &  9.7 & $<$38.6 &      & AB &    &    \\
     NGC4020 & 179.736150 &  30.412470 & 14.3 & -18.2 &  8.7 &  9.5 & $<$39.0 &      & B  &    &    \\
     NGC4102 & 181.597800 &  52.710970 & 17.1 & -19.5 &  8.8 & 10.1 & 40.4    & 1.00 & AB &    & L, F\\
     NGC4136 & 182.323800 &  29.927670 & 16.3 & -19.1 &  9.4 &  9.9 & 38.3    & 0.24 & AB &    &    \\
     NGC4151 & 182.635950 &  39.405790 & 11.2 & -19.2 &  9.0 & 10.1 & 41.1    & 1.00 & AB &    & Sy, F\\
     NGC4254 & 184.706550 &  14.416410 & 16.8 & -20.9 &  9.7 & 10.5 & $<$38.6 &      & A  &    &    \\
     NGC4258 & 184.740000 &  47.303880 &  7.6 & -20.9 &  9.6 & 10.7 & 39.8    & 0.99 & AB &    & Sy, F\\
     NGC4274 & 184.960650 &  29.614290 & 16.3 & -20.2 &  8.7 & 10.8 & $<$39.0 &      & B  &    & L, F\\
     NGC4303 & 185.478900 &   4.474180 & 17.6 & -21.2 &  9.8 & 10.6 & 39.2    & 0.53 & AB &    & Sy, F\\
     NGC4321 & 185.728200 &  15.821890 & 15.2 & -21.1 &  9.4 & 10.8 & 38.9    & 0.04 & AB &    &    \\
     NGC4414 & 186.612600 &  31.223390 & 17.7 & -20.8 &  9.5 & 10.9 & 39.4    & 0.46 & A  &    &    \\
    NGC4411A & 186.625050 &   8.871670 & 16.8 & -18.1 &  9.0 &  9.6 & $<$38.6 &      & B  &    &    \\
    NGC4411B & 186.696750 &   8.884500 & 16.8 & -18.3 &  9.2 &  9.5 & $<$38.6 &      & AB &    &    \\
     NGC4355 & 186.727650 &  -0.877560 & 29.0 & -18.8 &  8.7 &  9.7 & 38.8    & 0.81 & AB &    & Sy \\
     NGC4448 & 187.064400 &  28.620310 & 16.3 & -19.4 &  8.1 & 10.5 & $<$39.0 &      & B  &    &    \\
     NGC4450 & 187.123350 &  17.084990 & 16.8 & -20.6 &  8.5 & 10.8 & 40.1    & 0.99 & A  &    & L, F\\
     NGC4470 & 187.407450 &   7.823900 & 16.8 & -18.4 &  8.7 &  9.6 & $<$38.9 &      & A  &    &    \\
     NGC4492 & 187.748850 &   8.077690 & 16.8 & -18.1 &  7.6 &  9.5 & $<$38.6 &      & A  &    &    \\
     NGC4498 & 187.914900 &  16.852790 & 16.8 & -18.9 &  8.9 &  9.8 & $<$39.0 &      & AB &    &    \\
     NGC4501 & 187.996950 &  14.420140 & 16.8 & -21.6 &  9.3 & 11.0 & 39.4    & 0.43 & A  &    & Sy, F\\
     NGC4548 & 188.860200 &  14.496010 & 16.2 & -20.4 &  8.8 & 10.7 & 39.8    & 0.97 & B  &    & L  \\
     NGC4559 & 188.990400 &  27.959680 &  8.7 & -20.1 &  9.7 &  9.9 & 38.6    & 0.84 & AB &    &    \\
     NGC4579 & 189.431400 &  11.818000 & 16.8 & -21.0 &  8.8 & 11.0 & 41.1    & 1.00 & AB &    & Sy, F\\
     NGC4639 & 190.718250 &  13.257040 & 22.0 & -19.9 &  9.3 & 10.3 & 41.3    & 1.00 & AB &    & Sy \\
     NGC4647 & 190.885500 &  11.582210 & 16.8 & -19.3 &  8.7 & 10.0 & $<$38.9 &      & AB & M1 &    \\
     NGC4651 & 190.927650 &  16.393390 & 16.8 & -20.2 &  9.6 & 10.2 & $<$39.3 &      & A  & M0 &    \\
     NGC4654 & 190.985850 &  13.126570 & 16.8 & -20.7 &  9.5 & 10.4 & 39.0    & 0.57 & AB &    &    \\
     NGC4666 & 191.286000 &  -0.461860 & 15.7 & -20.3 &  9.3 & 10.5 & 39.1    & 0.55 & AB &    &    \\
     NGC4689 & 191.939850 &  13.762720 & 16.8 & -19.9 &  8.7 & 10.2 & $<$38.6 &      & A  &    &    \\
     NGC4713 & 192.491100 &   5.311290 & 15.6 & -18.9 &  9.4 &  9.6 & 38.5    & 0.18 & AB &    &    \\
     NGC4750 & 192.530100 &  72.874470 & 26.2 & -20.3 &  9.1 & 10.4 & 40.3    & 1.00 & A  &    & L  \\
     NGC4725 & 192.610950 &  25.500910 & 12.4 & -20.8 &  9.6 & 10.7 & 39.1    & 0.37 & AB &    & Sy \\
     NGC4736 & 192.720750 &  41.119990 &  4.4 & -19.7 &  8.3 & 10.4 & 38.7    & 0.64 & A  &    & Sy, F\\
     NGC4772 & 193.371600 &   2.168330 & 15.6 & -19.4 &  8.9 & 10.5 & 39.9    & 1.00 & A  &    & L, F\\
    UGC08041 & 193.802700 &   0.116680 & 23.0 & -18.8 &  9.4 &  9.8 & $<$38.9 &      & B  &    &    \\
     NGC4826 & 194.182500 &  21.681950 &  4.4 & -19.5 &  8.2 & 10.3 & $<$37.9 &      & A  &    & Sy, F\\
     NGC4900 & 195.163350 &   2.501010 & 15.6 & -19.2 &  9.1 &  9.8 & 38.5    & 0.09 & B  &    &    \\
     NGC4904 & 195.244350 &  -0.027520 & 23.0 & -19.6 &  9.3 & 10.1 & $<$39.2 &      & B  &    & F  \\
     NGC5033 & 198.364650 &  36.593710 & 18.5 & -21.3 & 10.0 & 10.6 & 41.2    & 1.00 & A  &    & Sy, F\\
     NGC5055 & 198.955500 &  42.029220 &  9.0 & -20.9 &  9.6 & 10.8 & 38.7    & 0.10 & A  &    &    \\
     NGC5068 & 199.728900 & -21.039050 &  9.0 & -19.7 &  9.3 & 10.1 & $<$37.9 &      & AB &    &    \\
     NGC5194 & 202.469550 &  47.195150 &  8.4 & -21.3 &  9.3 & 10.7 & 39.3    & 0.81 & A  & M2 & Sy, F\\
     NGC5240 & 203.979900 &  35.588250 & 35.8 & -19.3 &  9.0 & 10.0 & $<$38.6 &      & B  &    &    \\
     NGC5347 & 208.324200 &  33.490800 & 39.0 & -19.8 &  9.6 & 10.4 & 40.5    & 1.00 & B  &    & Sy, F\\
     NGC5350 & 208.340100 &  40.363940 & 30.9 & -20.6 &  9.7 & 10.8 & 39.8    & 0.82 & B  &    & F  \\
     NGC5457 & 210.802500 &  54.349060 &  7.0 & -20.9 &  9.9 & 10.4 & 38.1    & 0.03 & AB & M0 &    \\
     NGC5427 & 210.858750 &  -6.030590 & 27.0 & -20.4 &  9.8 & 10.3 & 39.4    & 0.29 & A  & M2 & Sy, F\\
    UGC09235 & 216.175650 &  35.267150 & 47.0 & -18.7 &  8.8 &  9.7 & $<$39.8 &      & B  &    &    \\
     NGC5656 & 217.606200 &  35.320910 & 48.8 & -20.5 &  9.5 & 10.4 & $<$38.0 &      &    &    &    \\
     NGC5678 & 218.023350 &  57.921420 & 27.9 & -20.4 &  9.4 & 10.4 & 39.2    & 0.22 & AB &    &    \\
     NGC5643 & 218.169600 & -44.174480 & 11.8 & -20.4 &  9.1 & 10.4 & 39.9    & 1.00 & AB &    & Sy \\
     NGC5728 & 220.599600 & -17.253010 & 24.8 & -20.4 &  9.1 & 10.7 & 40.7    & 1.00 & AB &    & Sy \\
     NGC5774 & 223.427100 &   3.582530 & 19.8 & -18.9 &  9.6 &  9.7 & $<$37.9 &      & AB & M1 &    \\
  ESO386-039 & 224.105550 & -37.600920 & 37.0 & -20.3 &  9.4 & 10.3 & $<$39.8 &      & B  & M1 &    \\
     NGC5954 & 233.646000 &  15.199390 & 25.5 & -20.0 &  9.2 & 10.2 & 38.9    & 0.49 & AB & M2 &    \\
     NGC5970 & 234.625050 &  12.186110 & 29.8 & -20.8 &  9.7 & 10.7 & $<$39.1 &      & B  &    &    \\
     NGC6500 & 268.999050 &  18.338290 & 46.3 & -20.9 &  9.8 & 10.6 & 40.6    & 1.00 & A  & M0 & F  \\
     NGC6643 & 274.943100 &  74.568390 & 20.6 & -20.6 &  9.4 & 10.4 & $<$38.5 &      & A  &    &    \\
     NGC6764 & 287.068200 &  50.933190 & 23.3 & -19.8 &  9.2 & 10.2 & 39.7    & 1.00 & B  &    &    \\
  ESO184-060 & 290.672250 & -54.585270 & 41.1 & -18.8 &  9.2 &  9.8 & $<$38.9 &      & AB &    &    \\
    UGC11466 & 295.744800 &  45.298110 & 18.1 & -18.9 &  9.9 &  9.8 & $<$39.0 &      &    &    &    \\
     NGC7314 & 338.941650 & -26.050500 & 15.9 & -20.1 &  9.2 & 10.2 & $<$38.2 &      & AB &    & Sy \\
     NGC7320 & 339.014250 &  33.948160 & 12.9 & -18.1 &  8.4 &  9.4 & 39.7    & 1.00 & A  &    &    \\
     NGC7331 & 339.267300 &  34.415620 & 14.7 & -21.6 &  9.9 & 11.1 & 39.0    & 0.03 & A  &    & L  \\
     NGC7424 & 344.326500 & -41.070640 &  7.9 & -19.0 &  9.4 &  9.6 & $<$37.3 &      & AB &    &    \\
     NGC7552 & 349.044900 & -42.584280 & 18.7 & -20.3 &  9.4 & 10.5 & 39.5    & 0.89 & B  &    &    \\
     NGC7582 & 349.596750 & -42.369990 & 18.7 & -20.5 &  9.4 & 10.6 & 40.6    & 1.00 & B  &    & Sy \\
      IC5325 & 352.180950 & -41.333390 & 18.7 & -19.7 &  8.8 & 10.0 & $<$38.7 &      & AB &    &    \\
      IC5332 & 353.614500 & -36.101450 &  9.9 & -18.8 &  9.4 &  9.8 & $<$37.2 &      & A  &    &    \\
     NGC7714 & 354.058800 &   2.155020 & 35.0 & -20.2 &  9.8 & 10.1 & 40.8    & 1.00 & B  & M2 & F  \\
     NGC7716 & 354.131100 &   0.297290 & 35.0 & -20.2 &  9.6 & 10.4 & $<$39.6 &      & AB &    &    \\
     NGC7727 & 354.974400 & -12.292840 & 26.9 & -20.9 &  8.4 & 11.1 & 38.6    & 0.00 & AB & M3 &    \\
     NGC7741 & 355.976850 &  26.075470 & 15.1 & -19.7 &  9.3 &  9.9 & $<$38.7 &      & B  &    &    \\

\enddata 

\tablecomments{HI mass, stellar mass, and X-ray luminosity are
  expressed as logarithms. Prob (probability) is the estimated
  likelihood that the detected nuclear X-ray source is associated with
  a SMBH rather than X-ray binary contamination. Bar has A for no bar,
  AB for weak or intermediate bar, and B for barred galaxies, taken
  from NED. Act (activity) has Sy for Seyfert 1 or 2, L for LINER, and
  F for FIRST radio point source (except NGC 6500 is marked F for flat
  spectrum radio source from NED).}

\end{deluxetable*}

\end{document}